\documentclass[preprint,aps,prx]{revtex4-2}
\usepackage[colorlinks=true, citecolor=blue, linkcolor=blue, urlcolor=blue]{hyperref}
\usepackage{graphicx}
\usepackage{amsmath}
\usepackage{amsfonts,amssymb} 
\usepackage{newtxmath} 
\usepackage{bm}
\allowdisplaybreaks

\begin{document}
	
	\title{Theory of Deterministic Photon-Loss Subspaces for Quantum Interferences}
	
	\author{Yadi Niu$^{1}$}
	\author{Haoyang Zhang$^{1,5}$}
	\author{Zihan Mo$^{1}$}
	\author{Nuo Wang$^{1}$}
	\author{Ying Gu$^{1,2,3,4,5,}$}
	\email[]{ygu@pku.edu.cn}
	
	\affiliation{$^1$State Key Laboratory of Artificial Microstructure and Mesoscopic Physics $\&$ Department of Physics, Peking University, Beijing 100871, China\\ 
	$^2$Frontiers Science Center for Nano-optoelectronics $\&$  Collaborative Innovation Center of Quantum Matter, Peking University, Beijing 100871, China\\
	$^3$Collaborative Innovation Center of Extreme Optics, Shanxi University, Taiyuan, Shanxi 030006, China\\
	$^4$Peking University Yangtze Delta Institute of Optoelectronics, Nantong 226010, China\\
	$^5$Hefei National Laboratory, Hefei 230088, China}

	\date{\today}
	
\begin{abstract}
Quantum coherence, quantum decoherence, and photon number reduction are coexistent in linear lossy optical systems.
However, how these three elements combine together to determine the evolution of the quantum light remains unclear.
Here, based on singular value decomposition (SVD), we propose the theory of deterministic photon-loss subspace (DPLS) for quantum interferences in lossy systems. 
By performing an SVD of scattering matrices with singular values either 0 or 1,  a series of completely lossy and lossless input modes are first defined.
According to $n_1,\ldots, n_i$ photons in the first,\ldots, $i$-th lossy modes, the Hilbert space of the input states can be decomposed into a set of orthogonal subspaces $\mathcal{H}_{(n_1,\cdots,n_i)}^{\text{in}}$, i.e., deterministic photon-loss subspaces (DPLSs).
When the concept of DPLS is established, the input state can be projected onto these DPLSs.
In each DPLS, the photons in lossy modes will be completely dissipated, while those in lossless modes experience a unitary evolution. 
The output state is a statistical mixture of the evolved outcomes of all projections, since decoherence is a concurrent process.
Then, based on the DPLS theory, we not only revisited Anti-HOM interference and the distillation of quantum states, but also demonstrate a robust W-state generation for various input states in a three-port lossy system with one-dimensional DPLSs. 
Through investigating the loss-induced subspace structure of the system, our general theory for analyzing quantum state evolution in lossy systems explicitly reveals the interplay among quantum coherence, quantum decoherence, and photon number reduction.
By engineering the loss, the constructed DPLSs can be used to precisely control quantum interferences in dissipative systems, with potential applications in quantum state preparation, quantum logic operations, and other quantum information processes.

\end{abstract}
	
	\maketitle
	
\section{Introduction}
\label{Introduction}
Loss, existing widely in optical systems, plays an unignorable role in quantum optics and quantum information processes. 
Generally, loss can surpress quantum interference, thereby reducing the fidelity and success probability of  entanglement generation \cite{1_matthews2011heralding,2_heilmann2015novel,3_konno2024logical,Liu2024OE,Liu2025APN} and quantum gate  \cite{4_ralph2002linear,5_hofmann2002quantum,6_liu2025quantum}.
Nevertheless, it may also give rise to some intriguing quantum phenomena.
During two-photon interferometry, loss can tailor the HOM interference, leading to  the HOM dip occurring at a shorter distance  \cite{11_klauck2019observation,12_zhou2022characterization,Zhang2024PRA} and a transition from bosonic coalescence to anti-coalescence \cite{13_vetlugin2022anti,16_ehrhardt2022observation,17_hong2024loss,18_vest2017anti}. 
By introducing loss to construct coherent perfect absorption systems,  the input state can be tuned from complete absorption to complete transmission \cite{24_huang2014coherent,25_roger2015coherent,26_roger2016coherent,27_altuzarra2017coherent,28_jeffers2019nonlocal,29_vetlugin2021coherent,30_hernandez2022quantum}, which can be further utilized to quantum state filtering \cite{31_hardal2019quantum,32_wang2025high} and preparation \cite{33_vetlugin2022deterministic,34_lai2024room}. 
Through engineering the loss, in parity-time (PT) and anti-PT systems, nontrivial quantum interferences have been found, such as loss-induced transparency \cite{19_qin2021quantum,20_longhi2018quantum}, the sudden vanishing and revival of entanglement \cite{21_fang2022entanglement}, entanglement filtering \cite{22_selim2025selective}, and non-orthogonal quantum state discrimination \cite{23_chen2022quantum}. 
Obviously, impacts of loss on quantum systems are diverse and subtle. 
However, owing to the existence of loss, the scattering matrix of lossy systems is non-unitary, thus it fails to preserve bosonic commutation relations, making it difficult to  deal with the evolution of the quantum light.

To analyze quantum interference in lossy systems described by non-unitary scattering matrices, many theoretical works have been proposed. 
In earlier literatures \cite{35_barnett1996field,36_barnett1998quantum}, to preserve the bosonic commmutation relation of output field operators,  the Langevin noise operator is introduced to describe the quantum fluctuation originated  from the material absorption, whereby measurable statistical properties of the output state can be obtained, but analytical solutions remain unavailable.
Recently, singular value decomposition (SVD)  emerges as a powerful tool to solve quantum state evolution in lossy systems \cite{37_knoll1999quantum,38_hernandez2022generalized, 39_tischler2018quantum}.
Based on the SVD results of the scattering matrix, by introducing $N$ ancilla bosonic modes and corresponding absorption matrix, the $N$-dimensional non-unitary scattering matrix can be embedded into a $2N$-dimensional unitary transformation \cite{37_knoll1999quantum,38_hernandez2022generalized}. Alternatively, by adding an ancilla bosonic mode for each lossy channel to form a lossless two-mode beam splitter,  the initial non-unitary matrix  is extended  to a higher dimensional quasi-unitary transformation \cite{39_tischler2018quantum}. 
Rather than merely giving statistical average, the above SVD-based methods enable analytical solutions of the output quantum states. 
However, the extra degrees of freedom in Refs. \cite{37_knoll1999quantum,38_hernandez2022generalized, 39_tischler2018quantum}  hinder a clear elucidation of the intrinsic interplay among quantum interference, photon number reduction, and quantum decoherence. It is therefore highly desirable to establish a theoretical framework that inherently incorporates their joint effects, enabling the solution of quantum state evolution in lossy systems.

\begin{figure*}[t]
	\includegraphics[width=1\textwidth]{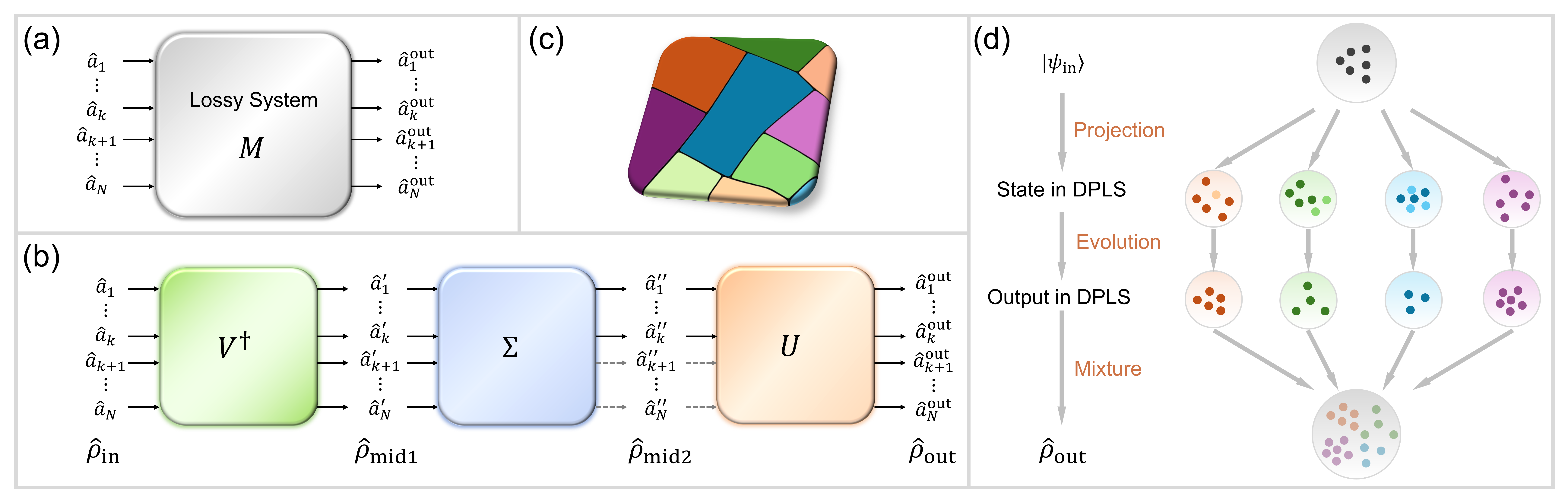}
	\caption{\label{fig:1} Schematic of DPLS theory. 
	(a) Lossy linear optical system with $N$ input and $N$ output modes, whose scattering matrix is $M$. According to SVD, the operation of $M$ on the input state can be equivalently described by (b) a sequential action of three processes: unitary transformation $V^{\dagger}$, non-unitary transformation $\Sigma$ and unitary transformation $U$. 
	(c) Schematic of DPLS decomposition. DPLSs are defined by photon number in the completely lossy modes, and are orthogonal to each other. 
	(d) Quantum state evolution based on DPLS theory. 
	The input state $\vert\psi_{\text{in}}\rangle$ is first projected into different DPLSs, shown by the colored particles in each large circle. Light (dark)-colored particles denote photons in lossy (lossless) modes. Each projection then experiences deterministic photon loss and unitary evolution, while the final output state is the statistical mixture of the evolved projections.}  
\end{figure*}

Here, based on SVD, we propose the theory of deterministic photon-loss subspace (DPLS) to solve the above problem in linear lossy optical systems in Fig. \ref{fig:1}.
We begin by constructing DPLSs.
By performing an SVD of a scattering matrix with singular values either 0 or 1 in Fig. \ref{fig:1} (b), a series of input completely lossy and lossless modes are first defined respectively. 
According to $n_1,\ldots, n_i$ photons in the first,\ldots, $i$-th lossy modes, the Hilbert space of the input states can be decomposed into a set of orthogonal loss-induced subspaces $\mathcal{H}_{(n_1,\cdots,n_i)}^{\text{in}}$, i.e., deterministic photon-loss subspaces (DPLSs), as shown in Fig. \ref{fig:1} (c). 
The dimension of each DPLS is determined by the number of input photons in lossless modes. Once the concept of DPLS is established, the input state can be projected onto these DPLSs [Fig. \ref{fig:1} (d)]. In each DPLS,  photons in lossy modes (light-colored particles) will be lost, while photons in lossless modes (dark-colored particles) experience a unitary evolution, where quantum coherence is preserved. Finally, the output state is a statistical mixture of the evolved outcomes of all projections, since decoherence is a concurrent process. 
	 	
Based on the DPLS theory, we have reinvestigated  the anti-HOM interference \cite{13_vetlugin2022anti,16_ehrhardt2022observation,17_hong2024loss,18_vest2017anti} and  quantum state distillation \cite{22_selim2025selective,31_hardal2019quantum} from more types of input states.
Then, we demonstrate a robust W-state generation for various input states in a three-port lossy system. 
By constructing one-dimensional DPLSs, the projections of different input states are identical in the same DPLS.  As a result, all the output states include W-state component but with different probabilities. This robust generation of $W$-state shows the capability of utilizing DPLS to prepare quantum states in lossy systems.
		
Through studying the DPLSs determined by SVD of the scattering matrix, our work establishes a general theory for analyzing state evolution in lossy systems, explicitly clarifying the interplay among quantum coherence, photon number reduction and quantum decoherence. By engineering the loss, the constructed DPLSs offer precise control over quantum interferences in dissipative systems, paving the way to quantum state preparation, quantum logic operations, and other quantum information processes. 
	
\newpage
	
\section{Theory}
For an $N$-mode linear lossless optical system, the transformation between the input operators $\hat{a}_j$ and output operators $\hat{a}_j^{\text{out}}$ can be described by the unitary matrix $M$ as
\begin{equation} 
	(\hat{a}_1^{\text{out}},\hat{a}_2^{\text{out}},\cdots,\hat{a}_N^{\text{out}})^T = M (\hat{a}_1,\hat{a}_2,\cdots,\hat{a}_N)^T.
	\label{eq1}
\end{equation}
However, for a lossy system where energy is not conserved, the scattering matrix $M$ is non-unitary, which can't preserve the bosonic commutation relations of the output operators.
To treat quantum state evolution described by non-unitary scattering matrices, in earlier works, Langevin noise operators \cite{35_barnett1996field,36_barnett1998quantum} are introduced. These operators model the noise sources in lossy systems, thereby preserve the bosonic commutation relations. Measurable statistical properties of the output states can be obtained using Langevin noise operators, while analytical form of the output state remains unavailable. Recently, SVD serves as a powerful tool to solve the problem \cite{37_knoll1999quantum,38_hernandez2022generalized,39_tischler2018quantum}, enabling analytical forms of output states. Based on the SVD results of the scattering matrix, ancilla modes can be introduced in two ways. One assigns an ancilla mode to each system mode and introduces an absorption matrix for the system’s nonunitary scattering matrix \cite{37_knoll1999quantum,38_hernandez2022generalized}. The other couples each lossy system mode to an ancilla mode through a lossless two-mode beam splitter \cite{39_tischler2018quantum}. These SVD-based methods embed the original non-unitary matrix into a higher-dimensional unitary transformation, thus providing analytical solutions of the output quantum states rather than merely giving statistical averages.  

However, in the aforementioned approaches, the influences of photon loss and quantum decoherence on quantum interference can only be observed by tracing out the introduced extra degrees of freedom, which hinders the clear elucidation of the intrinsic interplay among these three processes.
To overcome this drawback, we propose the theory of DPLS for quantum interference in lossy systems based on SVD. The input Hilbert space is decomposed into several DPLSs according to the SVD of the corresponding scattering matrix. Quantum interference is then analyzed by projecting the input states onto different DPLSs and investigating the evolution of projections, while extra degrees of freedom are unnecessary. More importantly, the interplay among quantum coherence, quantum decoherence, and photon number reduction is clearly revealed by DPLSs of the system, as will be shown in this section.
	
To clarify the DPLS theory, this section is organized as follows: in subsection A, the completely lossy and lossless input modes are defined according to the SVD results of the scattering matrix, which is a preliminary step towards constructing DPLSs of the system. In subsection B, the input Hilbert space is decomposed into a series of DPLSs based on the photon number in lossy modes. Basic properties of DPLSs, including their dimensions and the fact that they form a direct-sum decomposition of the input Hilbert space, are also discussed there. In subsection C, quantum state evolution is analyzed under the framework of DPLS theory. In subsection D, we take a $4\times 4$ non-unitary matrix as an example to illustrate how to apply DPLS theory.

\subsection{Completely lossy and lossless input modes}
As a prerequisite for DPLS construction, we first define completely lossy and lossless input modes through performing an SVD of the scattering matrix of the system. For a lossy system shown in Fig. \ref{eq1} (a), SVD of the scattering matrix $M$ gives: $M = U\Sigma V^{\dagger}$. Here $U$ and $V^{\dagger}$ are unitary matrices, and $\Sigma$ is a diagonal matrix whose diagonal elements are singular values of $M$. According to SVD, the quantum state evolution in the lossy system can be effectively decomposed into three steps shown in Fig. \ref{fig:1} (b): the input field $\hat{a}_i$ first experiences unitary evolution described by $V^\dagger$ to get $\hat{a}_i'$, then $\hat{a}_i'$ undergoes non-unitary transformation $\Sigma$ to get $\hat{a}_i''$, finally, $\hat{a}_i''$ experiences unitary evolution described by $U$ to get the output field $\hat{a}_i^{\text{out}}$. Since loss only occurs in the second step, completely lossy and lossless input modes can be defined accordingly by the input $(\hat{a}_i')$-output $(\hat{a}_i'')$ relation of $\Sigma$. Consider that the singular values of the scattering matrix are either 0 or 1, i.e., the diagonal matrix $\Sigma$ is 
\begin{equation}
	\Sigma=\text{diag}(1_1,\cdots,1_k,0_{k+1},\cdots,0_N).
	\label{Sigma}
\end{equation}
The first $k$ diagonal elements of $\Sigma$ are $\sigma=1$, meaning that the photons in modes $\hat{a}_i'$ $(1\le i\le k)$ undergo lossless evolution and are transferred to modes $\hat{a}_i''$ $(1\le i\le k)$, thus modes $\hat{a}_i'$ $(1\le i\le k)$ are identified as lossless modes. The last $(N-k)$ diagonal elements are $\sigma=0$, implying that the photons in modes $\hat{a}_i'$ $(k+1\le i\le N)$ are completely dissipated and modes $\hat{a}_i''$ $(k+1\le i\le N)$ are in vacuum states, so modes $\hat{a}_i'$ $(k+1\le i\le N)$ are identified as completely lossy modes (See Appendix \ref{Appendix Sigma}). 
	
By mapping the lossless modes to the input ports of the system through unitary transformation $V$, lossless input modes can be defined as
\begin{equation}
	\hat{b}_j^{\text{nls} \dagger}=\sum_{i=1}^{N}v_{ij} \hat{a}_i^\dagger, 1\le j\le k,
	\label{lossless input modes}
\end{equation}
where $v_{ij}$ are elements of matrix $V$ at the $i$-th rwo and the $j$-th column. The input photons in modes $\hat{b}_j^{\text{nls} \dagger} (1\le j\le k)$ will be transformed into modes $\hat{a}_j'^{\dagger} (1\le j\le k)$, thus will be perfectly transmitted without loss. Similarly, we can also define completely lossy input modes by mapping $\hat{a}_j'^{\dagger} (k+1\le j\le N)$ as 
\begin{equation}
	\hat{b}_j^{\text{ls} \dagger}=\sum_{i=1}^{N}v_{ij} \hat{a}_i^\dagger, k+1\le j\le N.
	\label{lossy input modes}
\end{equation}
The input photons in modes $\hat{b}_j^{\text{nls} \dagger} (k+1\le j\le N)$ will be transformed into modes $\hat{a}_j'^{\dagger} (k+1\le j\le N)$, thus will be completely dissipated. It should be noted that although $\hat{a}_j'$, $\hat{b}_j^{\text{ls}}$ and $\hat{b}_j^{\text{nls}}$ are all obtained by transforming $\hat{a}_j$ through $V^{\dagger}$, they are physically somewhat different: $\hat{b}_j^{\text{ls}}$ and $\hat{b}_j^{\text{nls}}$ are defined at the input ports of the entire system, while $\hat{a}_j'$ are defined at the  output ports of matrix $V^{\dag}$. Actually, singular values 0 and 1 corresponds to CPA modes and coherent perfect transmission (CPT) modes respectively. The existing studies about CPA are mainly limited to systems with two input and two output modes \cite{24_huang2014coherent,25_roger2015coherent,26_roger2016coherent,27_altuzarra2017coherent,28_jeffers2019nonlocal,29_vetlugin2021coherent,31_hardal2019quantum}, or systems with reciprocity \cite{30_hernandez2022quantum}. By contrast, the DPLS theory is more general, which can be applied to multimode systems whose scattering matrices have singular values equal to either 0 or 1.
	
\subsection{DPLS decomposition and their properties}

After obtaining modes $\hat{b}_j^{\text{nls} \dagger}$ and $\hat{b}_j^{\text{ls} \dagger}$, a new set of basis states of the input Hilbert space are defined according to the input photon number in these modes. These basis states are then regrouped into different subsets to decompose the input Hilbert space into a direct sum of subspaces. In this subsection, we illustrate the definitions and properties of those basis states and the corresponding subspaces. In subsection \ref{evolution in DPLS}, the reason for such a subspace decomposition is elucidated.
	
Consider the number of input photons in $\hat{b}_1^{\text{nls} \dagger},...,\hat{b}_k^{\text{nls} \dagger}$ is $(m_1,...,m_k)$, denoted as vector $\vec{m}\in \mathbb{N}^k$, and the number in $\hat{b}_{k+1}^{\text{ls} \dagger},...,\hat{b}_N^{\text{ls} \dagger}$ is $(n_{k+1},...,n_N)$, denoted as vector $\vec{n}\in \mathbb{N}^{N-k}$, the corresponding quantum state can be written as 
\begin{equation}
	\vert\varphi_{\vec{m}\vec{n}}\rangle=\prod_{j=1}^{k}\frac{1}{\sqrt{m_j!}}\left(\hat{b}_j^{\text{nls}\dagger}\right)^{m_j} \prod_{j'=k+1}^{N}\frac{1}{\sqrt{n_{j'}!}}\left(\hat{b}_{j'}^{\text{ls}\dagger}\right)^{n_{j'}}\vert0\rangle.
	\label{basis state}
\end{equation}
Here, the $N$-mode vacuum state $\vert0...0\rangle$ is simplified as $\vert0\rangle$. These states are actually obtained by transforming the Fock states $\vert m_1\ldots m_k n_{k+1}\ldots n_N\rangle$ using unitary matrix $V$, therefore they are orthogonal to each other, and can be regarded as basis states of the input Hilbert space. Basis states in Eq. (\ref{basis state}) with identical photon number distribution $\vec{n}$ in modes $\hat{b}_{j}^{\text{ls} \dagger}$ span a subspace 
\begin{equation}
	\mathcal{H}_{\vec{n}}^{\text{in}}=\text{Span}\left\{\vert\varphi_{\vec{m}\vec{n}}\rangle,\forall \vec{m}\in\mathbb{N}^k\right\},
	\label{DPLS n}
\end{equation}
while basis states with different $\vec{n}$ belong to different subspaces, as shown in Fig. \ref{fig:2}. We name this set of orthogonal subspaces as DPLSs. In Appendix \ref{Appendix Sigma} it's proved that quantum states in $\mathcal{H}_{\vec{n}}^{\text{in}}$ will dissipate all the photons initially in  $\hat{b}_{k+1}^{\text{ls}\dagger},...\hat{b}_{N}^{\text{ls}\dagger}$ deterministically, while transmit all photons in  $\hat{b}_{1}^{\text{nls}\dagger},...\hat{b}_{k}^{\text{nls}\dagger}$. During this process, quantum coherence is maintained. 	
Particularly, when there's no photon distributed in any $\hat{b}_j^{\text{ls}\dagger}$ mode, a special DPLS called lossless subspace is defined as
\begin{equation}
	\mathcal{H}_{\vec{0}}^{\text{in}}=\text{Span}\left\{\prod_{j=1}^{k} \frac{1}{\sqrt{m_j!}}\left(\hat{b}_j^{\text{nls}\dagger}\right)^{m_j}\vert 0\rangle, \forall m_j\ge 0\right\}.
	\label{lossless subspace}
\end{equation}
Quantum states in this subspace are immune to loss even in the dissipative system. 
	
\begin{figure*}[t]
	\includegraphics[width=0.8\textwidth]{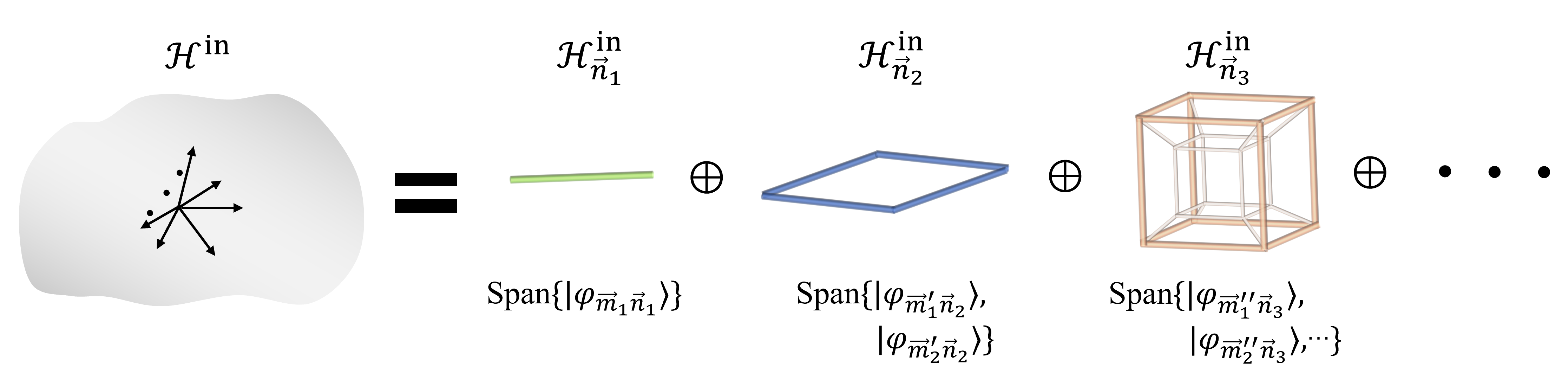}
	\caption{\label{fig:2} Schematic of decomposing the input Hilbert space into DPLSs $\mathcal{H}_{\vec{n}_1}^{\text{in}}$, $\mathcal{H}_{\vec{n}_2}^{\text{in}}$, $\mathcal{H}_{\vec{n}_3}^{\text{in}}$, $\cdots$. 
	Basis states in the same DPLS have identical photon number distribution $\vec{n}$, while the dimension of each DPLS is determined by the photon number distributions $\vec{m}$. All DPLSs form a direct sum decomposition of the Hilbert space $\mathcal{H}^{\text{in}}$.
	} 
\end{figure*}
	
Although DPLSs are defined through the number of photons in the completely lossy modes $\hat{b}_j^{\text{ls}\dagger}$, their dimensions are determined by the the photon number in the lossless modes $\hat{b}_j^{\text{nls}\dagger}$. From Eq. (\ref{DPLS n}), $\mathcal{H}_{\vec{n}}^{\text{in}}$ contains basis states with different $\vec{m}$, i.e., different distributions of input photons in the lossless modes. The number of possible $\vec{m}$ equals to the number of basis states in $\mathcal{H}_{\vec{n}}^{\text{in}}$, thereby determining the dimension of $\mathcal{H}_{\vec{n}}^{\text{in}}$, as shown in Fig. \ref{fig:2}. For example, if the input photon number is unlimited (such as continuous variable input states), then the photon number distributions in $\hat{b}_j^{\text{nls}\dagger}$ are also unlimited, corresponding to the infinite-dimensional DPLSs. If the total number of input photons is $N_p$, then basis states of $\mathcal{H}_{\vec{n}}^{\text{in}}$ with photon number distributions $\vec{m}=(m_1,...,m_k)$ and $\vec{n}=(n_{k+1},...,n_N)$ satisfy $N_t=\sum_{i=1}^{k}m_i=N_p-\sum_{i=k+1}^{N}n_i$, so the dimension of $\mathcal{H}_{\vec{n}}^{\text{in}}$ is 
\begin{equation}
	{N_t+k-1 \choose N_t}, k\ge 1,
	\label{DPLS dimension}
\end{equation}
which is exactly the number of possible distributions of $N_t$ photons among $k$ lossless modes. Since the number of photons in lossless input modes determines the number of possible $\vec{m}$, it determines the dimension of each DPLS. Another important property of DPLSs in fig. \ref{fig:2} is that they form a direct sum decomposition of the input Hilbert space, which insures that any input state can be uniquely superposed by states in each DPLS. This property is verified by proving that all DPLSs span the whole Hilbert space, and these DPLSs are independent to each other. Detailed proof is in Appendix \ref{Direct sum decomposition of Hilbert space}.

\subsection{DPLS-based analysis of quantum state evolution}\label{evolution in DPLS}
Once the concept of DPLS is established, we can use these DPLSs to analyze the quantum state evolution in lossy systems. The input state is first projected onto each DPLS. Then each projection dissipates all photons in modes $\hat{b}_j^{\text{ls}\dagger}$, while photons in modes $\hat{b}_j^{\text{nls}\dagger}$ experience unitary evolution. Quantum coherence is preserved for each projection during the above processes. The final output is the statistical mixture of all the evolved projections, since quantum decoherence simultaneously occurs. In this subsection, these processes are discussed in detail.

Using the basis states in Eq. (\ref{basis state}), the projector $\hat{P}_{\vec{n}}$ onto the DPLS  $\mathcal{H}_{\vec{n}}^{\text{in}}$ is
\begin{equation}
	\hat{P}_{\vec{n}}=\sum_{\vec{m}}\vert\varphi_{\vec{m}\vec{n}}\rangle\langle \varphi_{\vec{m}\vec{n}}\vert,\sum_{\vec{n}}\hat{P}_{\vec{n}}=\bm{I}.
	\label{projector Pn}
\end{equation}
Therefore, the projection of the input state $\vert\psi_{\text{in}}\rangle$ onto DPLS $\mathcal{H}_{\vec{n}}^{\text{in}}$ is
\begin{equation}
	\vert \psi_{\vec{n}}\rangle=\frac{1}{\sqrt{P_{\vec{n}}}}\sum_{\vec{m}}D_{\vec{m}\vec{n}}\vert\varphi_{\vec{m}\vec{n}}\rangle,
\end{equation}
where $D_{\vec{m}\vec{n}}=\langle \varphi_{\vec{m}\vec{n}}\vert\psi_{\text{in}}\rangle, P_{\vec{n}}=\langle\psi_{\text{in}}\vert\hat{P}_{\vec{n}}\vert\psi_{\text{in}}\rangle=\sum_{\vec{m}}\vert D_{\vec{m}\vec{n}}\vert^2$.
Next, the evolution of projection $\vert\psi_{\vec{n}}\rangle$ is considered. In the Schr\"{o}dinger picture, the quantum state transformation characterized by unitary matrix $V^{\dagger}$ can be described by a unitary operator
\begin{equation}
	\hat{S}_{V^{\dagger}}=\exp\left[-i(\hat{\bm{a}}^{\dagger})^T\phi_{V^{\dagger}}\hat{\bm{a}}\right]
	\label{definition of SV1}.
\end{equation}
Here, $\phi_{V^{\dagger}}=i\ln V^{\dagger}, \hat{\bm{a}}=\left(\hat{a}_1,\cdots,\hat{a}_N\right)^T
\label{definition of SV2}$.
Using operator $\hat{S}_{V^{\dagger}}$, Eq. (\ref{lossless input modes}) and (\ref{lossy input modes}) are equivalent to
\begin{equation}
	\hat{b}_j = \hat{S}_{V^{\dagger}}^{\dagger} \hat{a} _j \hat{S}_{V^\dagger}
	\label{ab transformaiton}.
\end{equation}
This equation is applicable to both $\hat{b}_j^{\text{nls}}$ and $\hat{b}_j^{\text{ls}}$, so the superscripts have been eliminated . 

Using Eq. (\ref{definition of SV1}) and Eq. (\ref{ab transformaiton}), $\vert\psi_{\vec{n}}\rangle$ is transformed into
\begin{equation}
	\begin{aligned}
		\vert\psi_{\vec{n},\text{mid1}}\rangle & =\hat{S}_{V^{\dagger}}\vert\psi_{\vec{n}}\rangle=\frac{1}{\sqrt{P_{\vec{n}}}}\sum_{\vec{m}} D_{\vec{m}\vec{n}}\hat{S}_{V^{\dagger}}\vert\varphi_{\vec{m}\vec{n}}\rangle\\
		& = \frac{1}{\sqrt{P_{\vec n}}}\sum_{\vec{m}} D_{\vec{m}\vec{n}} \,\bigl\vert m_1 m_2 \cdots m_k n_{k+1}\cdots n_N\bigr\rangle,
	\end{aligned}
	\label{S operates on the input state}
\end{equation}
i.e., the projection $\vert\psi_{\vec{n}}\rangle$ evolves into state $\vert\psi_{\vec{n},\text{mid1}}\rangle$, which is the superposition of Fock states that have the same photon number distribution $\vec{n}$ in lossy modes and different photon number distributions $\vec{m}$ in lossless modes. In Appendix \ref{Appendix Sigma} It's proved that after the operation of non-unitary transformation $\Sigma$, quantum coherence is maintained for $\vert\psi_{\vec{n},\text{mid1}}\rangle$, while all the photons distributed in modes $\hat{a}_j'(k+1\le j\le N)$ are dissipated, as shown in Fig. \ref{fig:1} d, so the output state of  $\vert\psi_{\vec{n},\text{mid1}}\rangle$ after $\Sigma$ is 
\begin{equation}
	\begin{aligned}
		\vert\psi_{\vec{n},\text{mid2}}\rangle & = \frac{1}{\sqrt{P_{\vec{n}}}}D_{\vec{m}\vec{n}} \,\bigl\vert m_1 m_2 \cdots m_k 0_{k+1}\cdots 0_N\bigr\rangle \\
		& = \prod_{j=k+1}^{N}\frac{\left(\hat{a}_j\right)^{n_j}}{\sqrt{n_j!}} \left\vert\psi_{\vec{n},\text{mid1}}\right\rangle.
	\end{aligned}
\end{equation}
Therefore, the photon loss process that transfers  $\left\vert\psi_{\vec{n},\text{mid1}}\right\rangle$ into $\left\vert\psi_{\vec{n},\text{mid2}}\right\rangle$ can be described by annihilation operators $\hat{a}_j$. The output state of projection $\vert\psi_{\vec{n}}\rangle$ is further obtained by applying the unitary transformation $U$ to $\left\vert\psi_{\vec{n},\text{mid2}}\right\rangle$ as
\begin{equation}
	\begin{aligned}
		\vert\psi_{\vec{n},\text{out}}\rangle & =\hat{S}_U \left\vert\psi_{\vec{n},\text{mid2}}\right\rangle \\
		& = \hat{S}_U\prod_{j=k+1}^{N}\frac{\left(\hat{a}_j\right)^{n_j}}{\sqrt{n_j!}}\hat{S}_{V^{\dagger}}\vert\psi_{\vec{n}}\rangle.
	\end{aligned}
\end{equation}
Consequently, we can define an operator to describe the photon loss process and unitary transformation in DPLS $\mathcal{H}_{\vec{n}}^{\text{in}}$ as
\begin{equation}
	\begin{aligned}
		\hat{M}_{\vec{n}} & = \hat{S}_U\prod_{j=k+1}^{N}\frac{\left(\hat{a}_j\right)^{n_j}}{\sqrt{n_j!}}\hat{S}_{V^{\dagger}}\\
		& = \hat{S}_U\hat{S}_{V^{\dagger}}\prod_{j=k+1}^{N}\frac{\left(\hat{b}_j^{\text{ls}}\right)^{n_j}}{\sqrt{n_j!}}.
	\end{aligned}
	\label{operator M}
\end{equation}
In Eq. (\ref{operator M}), $\hat{S}_{V^\dagger}^\dagger \hat{S}_{V^\dagger}=\hat{S}_{V^\dagger}\hat{S}_{V^\dagger}^\dagger=\bm{I}$ as well as Eq. (\ref{ab transformaiton}) are used. It's obvious from operator $\hat{M}_{\vec{n}}$ that the projections will dissipate all photons in modes $\hat{b}_j^{\text{ls}}$, the remaining part then experience unitary transformation $\hat{S}_U\hat{S}_{V^{\dagger}}$. More importantly, quantum coherence of the projection in $\mathcal{H}_{\vec{n}}^{\text{in}}$ is completely preserved during the above processes.

In the previous part, only the evolution of projection in each DPLS is considered. Now we investigate the mapping that transforms the superposition of projections into the output state. The input state can be decomposed into projections in different DPLSs as 
\begin{equation}
	\left\vert \psi_{\mathrm{in}}\right\rangle=\sum_{\vec{n}}\sqrt{P_{\vec{n}}}\left\vert\psi_{\vec{n}}\right\rangle.
\end{equation}
Using Eq. (\ref{S operates on the input state}), after the operation of $V^{\dagger}$, $\left|\psi_{\mathrm{in}}\right\rangle$ is then turned into
\begin{equation}
	\begin{aligned}
		\left|\psi_{\mathrm{mid1}}\right\rangle
		& = S_{V^{\dagger}} \left|\psi_{\text{in}}\right\rangle\\
		& = \sum_{\vec{n}} \sqrt{P_{\vec{n}}} \left|\psi_{\vec{n},\mathrm{mid1}}\right\rangle,
	\end{aligned}
	\label{psi mid1}
\end{equation}
which is a superposition of states $\left|\psi_{\vec{n},\mathrm{mid1}}\right\rangle$ with different photon number distributions $\vec{n}$ in lossy modes. In Appendix \ref{Appendix Sigma} it's proved that this superposed state will also dissipate all photons in lossy modes. However, different from the evolution in each DPLS, after the operation of $\Sigma$, this superposed state becomes a statistical mixture of $\vert\psi_{\vec{n},\text{mid2}}\rangle$ with different $\vec{n}$
\begin{equation}
	\begin{aligned}
		\hat{\rho}_{\mathrm{mid2}}
		& =
		\sum_{\vec{n}} P_{\vec{n}} 
		\left[
		\sum_{\vec{m}} \frac{D_{\vec{m}\vec{n}}}{\sqrt{P_{\vec{n}}}}
		\left|m_1 m_2 \cdots m_k\,0\cdots 0\right\rangle
		\right]\\
		&\qquad\qquad\left[
		\sum_{\vec{m}^{\,\prime}} \frac{D_{\vec{m}^{\,\prime}\vec{n}}^{\,*}}{\sqrt{P_{\vec{n}}}}
		\left\langle m_1^{\,\prime} m_2^{\,\prime} \cdots m_k^{\,\prime}\,0\cdots 0\right|
		\right]
		\\[4pt]
		& = \sum_{\vec{n}} P_{\vec{n}}
		\left|\psi_{\vec{n},\mathrm{mid2}}\right\rangle
		\left\langle\psi_{\vec{n},\mathrm{mid2}}\right|.
	\end{aligned}
\end{equation}
After the operation of $U$, $\hat{\rho}_{\mathrm{mid2}}$ is turned into the final output state
\begin{equation}
	\begin{aligned}
		\hat{\rho}_{\mathrm{out}}
		& = S_U\hat{\rho}_{\mathrm{mid2}}S_{U^{\dagger}}\\
		& = \sum_{\vec{n}} P_{\vec{n}}\,
		\left|\psi_{\vec{n},\mathrm{out}}\right\rangle
		\left\langle\psi_{\vec{n},\mathrm{out}}\right|
		\\[4pt]
		& = \sum_{\vec{n}} P_{\vec{n}}\,
		\hat{M}_{\vec{n}}\left|\psi_{\vec{n}}\right\rangle
		\left\langle\psi_{\vec{n}}\right|\hat{M}_{\vec{n}}^{\dagger}.
		\label{entire output}
	\end{aligned}
\end{equation}
Therefore, the output of the system is just the probabilistic mixture of the evolved projections in different DPLSs. The reason for decomposing the input Hilbert space into such DPLSs is that quantum state in each DPLS maintains its quantum coherence even in the presence of photon loss, while quantum decoherence occurs for the superposition of states that are from different DPLSs, thereby clearly revealing the interplay among photon loss, quantum coherence and quantum decoherence.

Up to now, we have established the theoretical framework of DPLS. Although we mainly considered scattering matrices with singular values $\sigma\in \{0,1\}$ in previous sections, it's proved that our theory can also be applied to scattering matrices with singular values $\sigma\in [0,1]$. The main idea is that for each mode corresponding to a singular value $\sigma\neq0,1$, an ancilla mode is introduced, and the scattering matrix is enlarged in such a way that its new singular values satisfy $\sigma\in \{0,1\}$, therefore DPLSs can be identified to analyze quantum interference. Detailed information is in Appendix \ref{SV between 0 & 1}.
	
\subsection{Example}
Here, we consider a four-mode lossy system with a non-unitary scattering matrix to illustrate how DPLS theory sheds light on the interplay among photon loss, quantum coherence and quantum decoherence. The scattering matrix and the corresponding SVD is
\begin{equation*}
	\begin{aligned}
		M & =
		\begin{pmatrix}
			\frac{-1+2\sqrt{2}}{4\sqrt{3}} & \frac{1}{2} & \frac{2+\sqrt{2}}{4\sqrt{3}} & \frac{\sqrt{3}}{4} \\
			\frac{1+2\sqrt{2}}{4\sqrt{3}} & \frac{1}{2} & \frac{2-\sqrt{2}}{4\sqrt{3}} & -\frac{\sqrt{3}}{4} \\
			0 & 0 & 0 & 0\\
			-\frac{1}{2\sqrt{6}} & 0 & \frac{1}{2\sqrt{3}} & \frac{\sqrt{6}}{4}
		\end{pmatrix}\\
		& = U\Sigma V^{\dagger},
	\end{aligned}
\end{equation*}
where
\begin{equation*}
	U =
	\begin{pmatrix}
		\frac{1}{\sqrt{2}} & -\frac{1}{2} & -\frac{1}{\sqrt{6}} & -\frac{1}{2\sqrt{3}} \\
		\frac{1}{\sqrt{2}} & \frac{1}{2} & \frac{1}{\sqrt{6}} & \frac{1}{2\sqrt{3}} \\
		0 & 0 & \frac{1}{\sqrt{3}} & -\frac{\sqrt{2}}{\sqrt{3}} \\
		0 & -\frac{1}{\sqrt{2}} & \frac{1}{\sqrt{3}} & \frac{1}{\sqrt{6}}
	\end{pmatrix}, 
\end{equation*}
\begin{equation*}
	V^{\dagger} =
	\begin{pmatrix}
		\frac{1}{\sqrt{3}} & \frac{1}{\sqrt{2}} & \frac{1}{\sqrt{6}} & 0 \\
		\frac{1}{2\sqrt{3}} & 0 & -\frac{1}{\sqrt{6}} & -\frac{\sqrt{3}}{2} \\
		\frac{1}{\sqrt{3}} & -\frac{1}{\sqrt{2}} & \frac{1}{\sqrt{6}} & 0 \\
		\frac{1}{2} & 0 & -\frac{1}{\sqrt{2}} & \frac{1}{2}
	\end{pmatrix}, 
\end{equation*}
\begin{equation*}
	\Sigma =
	{\renewcommand{\arraystretch}{1.25}\setlength{\arraycolsep}{12pt}
		\begin{pmatrix}
			1 & 0 & 0 & 0 \\
			0 & 1 & 0 & 0 \\
			0 & 0 & 0 & 0 \\
			0 & 0 & 0 & 0
	\end{pmatrix}}.
\end{equation*}
According to the above SVD, it's obvious that $\hat{a}_{1,2}'^{\dagger}$ are lossless modes, corresponding to singular value 1, while $\hat{a}_{3,4}'^{\dagger}$ are completely lossy modes, corresponding to singular value 0, as shown in Fig. \ref{fig:1}(b). These four modes are then mapped to the completely lossy and lossless input modes
\begin{equation}
	\begin{aligned}
		\hat{b}_1^{\mathrm{nls}\,\dagger} &= \frac{\sqrt{2}\,\hat{a}_1^\dagger+\sqrt{3}\,\hat{a}_2^\dagger+\hat{a}_3^\dagger}{\sqrt{6}},\\
		\hat{b}_2^{\mathrm{nls}\,\dagger} &= \frac{\hat{a}_1^\dagger-\sqrt{2}\,\hat{a}_3^\dagger-3\,\hat{a}_4^\dagger}{2\sqrt{3}},\\
		\hat{b}_3^{\mathrm{ls}\,\dagger}  &= \frac{\sqrt{2}\,\hat{a}_1^\dagger-\sqrt{3}\,\hat{a}_2^\dagger+\hat{a}_3^\dagger}{\sqrt{6}},\\
		\hat{b}_4^{\mathrm{ls}\,\dagger}  &= \frac{\hat{a}_1^\dagger-\sqrt{2}\,\hat{a}_3^\dagger+\hat{a}_4^\dagger}{2}.
	\end{aligned}
\end{equation}
Since there are two lossless and two completely lossy input modes, the photon distribution $\vec{m}$ and $\vec{n}$ are both two-dimensional vectors: $\vec{m}=(m_1,m_2)$ and $\vec{n}=(n_1,n_2)$. Consider the two-photon input case. According to the photon number distributions in $\hat{b}_3^{\text{ls}\dagger}$ and $\hat{b}_4^{\text{ls}\dagger}$, there should be 6 DPLSs: $\mathcal{H}_{(0,0)}^{\text{in}},\mathcal{H}_{(1,0)}^{\text{in}},\mathcal{H}_{(0,1)}^{\text{in}},\mathcal{H}_{(2,0)}^{\text{in}},\mathcal{H}_{(0,2)}^{\text{in}},\mathcal{H}_{(1,1)}^{\text{in}}$. First is the lossless subspace
\begin{equation}
	\begin{aligned}
		\mathcal{H}_{(0,0)}^{\mathrm{in}}
		= \text{Span}\biggl\{ &
		\frac{1}{\sqrt{2}}\left(\hat{b}_1^{\mathrm{nls}\,\dagger}\right)^2 \vert 0\rangle,
		\frac{1}{\sqrt{2}}\left(\hat{b}_2^{\mathrm{nls}\,\dagger}\right)^2 \vert 0\rangle,\\
		& \hat{b}_1^{\mathrm{nls}\,\dagger}\hat{b}_2^{\mathrm{nls}\,\dagger}\vert 0\rangle
		\biggr\},
	\end{aligned}
\end{equation}
where there is no photon in $\hat{b}_3^{\text{ls}\dagger}$ and $\hat{b}_4^{\text{ls}\dagger}$. Basis states of this three-dimensional DPLS are 
\begin{equation*}
	\begin{aligned}
		\left|\varphi_{(2,0,0,0)}\right\rangle
		&= \frac{1}{\sqrt{2}}\left(\hat{b}_1^{\mathrm{nls}\,\dagger}\right)^2 \left|0\right\rangle \\
		&=\frac{1}{3} \left|2000\right\rangle + \frac{1}{2}\left|0200\right\rangle + \frac{1}{6}\left|0020\right\rangle \\
		&\quad+ \frac{\sqrt{3}}{3}\left|1100\right\rangle + \frac{\sqrt{6}}{6}\left|0110\right\rangle + \frac{1}{3}\left|1010\right\rangle,
		\\[6pt]
		\left|\varphi_{(0,2,0,0)}\right\rangle
		&= \frac{1}{\sqrt{2}}\left(\hat{b}_2^{\mathrm{nls}\,\dagger}\right)^2 \left|0\right\rangle \\
		&= \frac{1}{12}\left|2000\right\rangle + \frac{1}{6}\left|0020\right\rangle + \frac{3}{4}\left|0002\right\rangle \\
		& \quad- \frac{1}{6}\left|1010\right\rangle - \frac{\sqrt{2}}{4}\left|1001\right\rangle + \frac{1}{2}\left|0011\right\rangle,
		\\[6pt]
		\left|\varphi_{(1,1,0,0)}\right\rangle
		&= \hat{b}_1^{\mathrm{nls}\,\dagger}\hat{b}_2^{\mathrm{nls}\,\dagger}\left|0\right\rangle \\
		&= \frac{\sqrt{2}}{6}\left|2000\right\rangle - \frac{\sqrt{2}}{12}\left|1010\right\rangle - \frac{1}{2}\left|1001\right\rangle \\
		&\quad + \frac{\sqrt{6}}{12}\left|1100\right\rangle - \frac{\sqrt{3}}{6}\left|0110\right\rangle - \frac{\sqrt{6}}{4}\left|0101\right\rangle \\
		&\quad - \frac{\sqrt{2}}{6}\left|0020\right\rangle - \frac{\sqrt{2}}{4}\left|0011\right\rangle .
	\end{aligned}
\end{equation*}
Likewise, the other 5 DPLSs are respectively 
\begin{equation}
	\begin{aligned}
		\mathcal{H}_{(1,0)}^{\mathrm{in}} &= \text{Span}\left\{
		\hat{b}_1^{\mathrm{nls}\,\dagger}\hat{b}_3^{\mathrm{ls}\,\dagger}\left|0\right\rangle,\;
		\hat{b}_2^{\mathrm{nls}\,\dagger}\hat{b}_3^{\mathrm{ls}\,\dagger}\left|0\right\rangle
		\right\},\\[4pt]
		\mathcal{H}_{(0,1)}^{\mathrm{in}} &= \text{Span}\left\{
		\hat{b}_1^{\mathrm{nls}\,\dagger}\hat{b}_4^{\mathrm{ls}\,\dagger}\left|0\right\rangle,\;
		\hat{b}_2^{\mathrm{nls}\,\dagger}\hat{b}_4^{\mathrm{ls}\,\dagger}\left|0\right\rangle
		\right\},\\[4pt]
		\mathcal{H}_{(2,0)}^{\mathrm{in}} &= \text{Span}\left\{
		\frac{1}{\sqrt{2}}\left(\hat{b}_3^{\mathrm{ls}\,\dagger}\right)^2\left|0\right\rangle
		\right\},\\[4pt]
		\mathcal{H}_{(0,2)}^{\mathrm{in}} &= \text{Span}\left\{
		\frac{1}{\sqrt{2}}\left(\hat{b}_4^{\mathrm{ls}\,\dagger}\right)^2\left|0\right\rangle
		\right\},\\[4pt]
		\mathcal{H}_{(1,1)}^{\mathrm{in}} &= \text{Span}\left\{
		\hat{b}_3^{\mathrm{ls}\,\dagger}\hat{b}_4^{\mathrm{ls}\,\dagger}\left|0\right\rangle
		\right\}.
	\end{aligned}
\end{equation}
corresponding to $(1,0),(0,1),(2,0),(0,2),(1,1)$ in modes $\hat{b}_{3}^{\text{ls}\dagger}$ and $\hat{b}_{4}^{\text{ls}\dagger}$  respectively. These six DPLSs are orthogonal to each other, and their dimensions satisfy
\begin{equation}
	\sum_{i+j\le2}\dim\left(\mathcal{H}_{(i,j)}^{\text{in}}\right)=10
\end{equation}
which is exactly the dimension of the Hilbert space of input state, so they form a direct sum decomposition of the input state space.

\begin{figure*}[t]
	\includegraphics[width=0.8\textwidth]{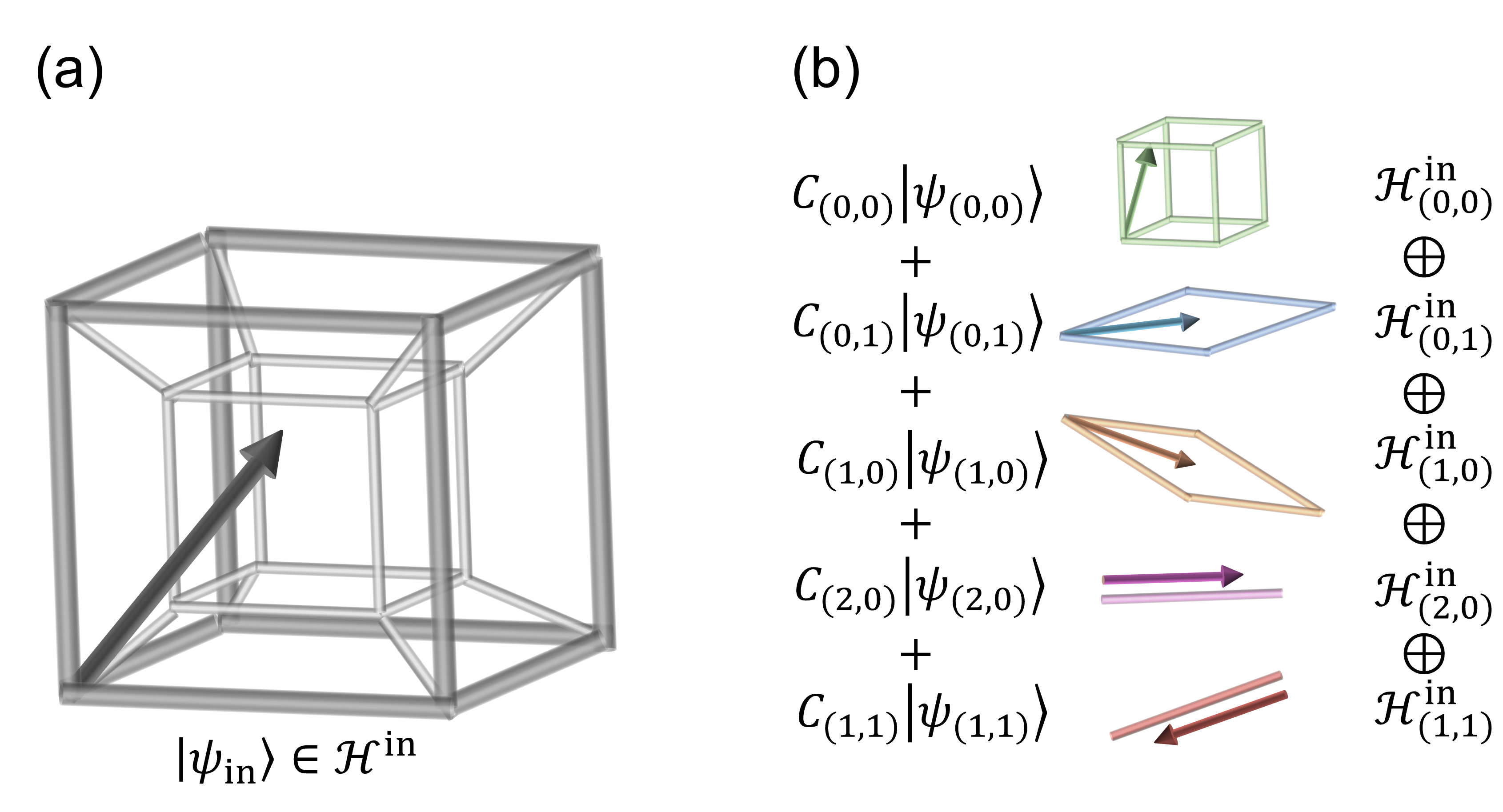}
	\caption{\label{fig:3} Schematic of the input state $\vert\psi_{\text{in}}\rangle=\vert1100\rangle$ projecting into different DPLSs. (a) Input state $\vert\psi_{\text{in}}\rangle$ in Hilbert space $\mathcal{H}^{\text{in}}$ can be decomposed into (b) projections in different $\mathcal{H}_{(i,j)}^{\text{in}}$ with corresponding possibilities $|C_{(0,0)}|^2=\frac{3}{8},\;|C_{(0,1)}|^2=\frac{1}{8},\;|C_{(1,0)}|^2=\frac{1}{24},\;|C_{(2,0)}|^2=\frac{1}{3}, $ and $ \;|C_{(1,1)}|^2=\frac{1}{8}$.}
\end{figure*}

Using the above DPLSs, quantum interference is analyzed in this lossy system. Consider the input two-photon state $\vert 1100\rangle$, which can be decomposed as
\begin{equation*}
	\begin{aligned}
		|1100\rangle
		&
		=\frac{\sqrt{6}}{4}|\psi_{(0,0)}\rangle
		+\frac{\sqrt{2}}{4}|\psi_{(0,1)}\rangle
		-\frac{\sqrt{6}}{12}|\psi_{(1,0)}\rangle\\
		&\quad-\frac{\sqrt{3}}{3}|\psi_{(2,0)}\rangle
		-\frac{\sqrt{2}}{4}|\psi_{(1,1)}\rangle\\
		&= \frac{\sqrt{6}}{4}\Bigg[
		\frac{1}{3}\hat b_{1}^{\mathrm{nls}\dagger}
		\hat b_{2}^{\mathrm{nls}\dagger}|0\rangle
		+\frac{2\sqrt{2}}{3}\,\frac{1}{\sqrt{2}}
		\Big(\hat b_{1}^{\mathrm{nls}\dagger}\Big)^{2}|0\rangle
		\Bigg] \\
		&\quad +\frac{\sqrt{2}}{4}
		\hat b_{1}^{\mathrm{nls}\dagger}
		\hat b_{4}^{\mathrm{ls}\dagger}|0\rangle
		-\frac{\sqrt{6}}{12}
		\hat b_{2}^{\mathrm{nls}\dagger}
		\hat b_{3}^{\mathrm{ls}\dagger}|0\rangle \\
		&\quad -\frac{\sqrt{3}}{3}\frac{1}{\sqrt{2}}
		\left(\hat b_{3}^{\mathrm{ls}\dagger}\right)^{2}|0\rangle
		-\frac{\sqrt{2}}{4}
		\hat b_{3}^{\mathrm{ls}\dagger}
		\hat b_{4}^{\mathrm{ls}\dagger}|0\rangle
	\end{aligned}
\end{equation*}
Here $\left|\psi_{(i,j)}\right\rangle$ represents the projection of $\vert\psi_{\text{in}}\rangle$ in DPLS $\mathcal{H}_{(i,j)}^{\text{in}}$, as shown in Fig. \ref{fig:3}. The corresponding probabilities are $
P_{(0,0)}=\frac{3}{8},P_{(0,1)}=\frac{1}{8},P_{(1,0)}=\frac{1}{24},P_{(2,0)}=\frac{1}{3},P_{(1,1)}=\frac{1}{8}$.

\begin{table*}[t]
	\caption{\label{table:1}
		Density matrix of the output optical field for the input state
		$\lvert 1100\rangle$.}
	\centering
	\small
	\setlength{\tabcolsep}{4pt}
	\renewcommand{\arraystretch}{1.15}
	
	\begin{ruledtabular}
		\begin{tabular}{c*{9}{c}}
			& $\langle 0000\rvert$ & $\langle 0001\rvert$ & $\langle 0100\rvert$ & $\langle 1000\rvert$ & $\langle 0101\rvert$ & $\langle 0200\rvert$ & $\langle 1001\rvert$ & $\langle 1100\rvert$ & $\langle 2000\rvert$ \\
			\hline
			$\lvert 0000\rangle$ & 0.4583 & 0 & 0 & 0 & 0 & 0 & 0 & 0 & 0 \\
			\hline
			$\lvert 0001\rangle$ & 0 & 0.02083 & -0.01473 & 0.01473 & 0 & 0 & 0 & 0 & 0 \\
			$\lvert 0100\rangle$ & 0 & -0.01473 & 0.07292 & 0.05208 & 0 & 0 & 0 & 0 & 0 \\
			$\lvert 1000\rangle$ & 0 & 0.01473 & 0.05208 & 0.07292 & 0 & 0 & 0 & 0 & 0 \\
			\hline
			$\lvert 0101\rangle$ & 0 & 0 & 0 & 0 & 0.01042 & -0.03988 & 0.01042 & -0.04167 & -0.01905 \\
			$\lvert 0200\rangle$ & 0 & 0 & 0 & 0 & -0.03988 & 0.1527 & -0.03988 & 0.1595 & 0.07292 \\
			$\lvert 1001\rangle$ & 0 & 0 & 0 & 0 & 0.01042 & -0.03988 & 0.01042 & -0.04167 & -0.01904 \\
			$\lvert 1100\rangle$ & 0 & 0 & 0 & 0 & -0.04167 & 0.1595 & -0.04167 & 0.1667 & 0.07618 \\
			$\lvert 2000\rangle$ & 0 & 0 & 0 & 0 & -0.01905 & 0.07292 & -0.01904 & 0.07618 & 0.03482 \\
		\end{tabular}
	\end{ruledtabular}
\end{table*}

Based on our theory, the projection in each DPLS will deterministically dissipate photons in completely lossy input modes, followed by unitary evolution. These processes are described by Eq. (\ref{operator M}). Therefore, the output of each projection is 
\begin{equation*}
	\begin{aligned}
		\left|\psi_{(0,0),\mathrm{out}}\right\rangle
		&= S_U\, S_{V^\dagger} \left|\psi_{(0,0)}\right\rangle \\
		&= \frac{1}{6}\Big[(2\sqrt{2}+1)\,|0200\rangle + (2\sqrt{2}-1)\,|2000\rangle \\
		&\quad\quad+ 4\,|1100\rangle - |1001\rangle - |0101\rangle\Big],\\[6pt]
		\left|\psi_{(0,1),\mathrm{out}}\right\rangle
		&= S_U\, \hat{a}_4\, S_{V^\dagger} \left|\psi_{(0,1)}\right\rangle \\
		&= \frac{|1000\rangle + |0100\rangle}{\sqrt{2}},\\[6pt]
		\left|\psi_{(1,0),\mathrm{out}}\right\rangle
		&= S_U\, \hat{a}_3\, S_{V^\dagger} \left|\psi_{(1,0)}\right\rangle \\
		&= -\frac{|1000\rangle - |0100\rangle + \sqrt{2}\,|0001\rangle}{2},\\[6pt]
		\left|\psi_{(2,0),\mathrm{out}}\right\rangle
		&= S_U \frac{\left(\hat{a}_3\right)^2}{\sqrt{2}}S_{V^\dagger}
		\left\vert \psi_{(2,0)}\right\rangle\\
		&= \vert 0000\rangle, \\[6pt]
		\left|\psi_{(1,1),\mathrm{out}}\right\rangle
		&= S_U \hat{a}_3\hat{a}_4S_{V^\dagger}
		\left\vert \psi_{(2,0)}\right\rangle\\
		&= |0000\rangle.
	\end{aligned}
\end{equation*}
These output components are still coherently superposed states or Fock states even after the photon loss processes. Since the superposition of projections experiences quantum decoherence, the final output optical field is the statistical mixture of $\left|\psi_{(i,j),\mathrm{out}}\right\rangle$, i.e.,
\begin{equation}
	\begin{aligned}
		\hat{\rho}_{\mathrm{out}}
		&= \frac{3}{8}\,|\psi_{(0,0),\mathrm{out}}\rangle\langle\psi_{(0,0),\mathrm{out}}|
		+ \frac{1}{8}\,|\psi_{(0,1),\mathrm{out}}\rangle\langle\psi_{(0,1),\mathrm{out}}| \\
		&\quad + \frac{1}{24}\,|\psi_{(1,0),\mathrm{out}}\rangle\langle\psi_{(1,0),\mathrm{out}}|
		+ \frac{1}{3}\,|\psi_{(2,0),\mathrm{out}}\rangle\langle\psi_{(2,0),\mathrm{out}}| \\
		&\quad + \frac{1}{8}\,|\psi_{(1,1),\mathrm{out}}\rangle\langle\psi_{(1,1),\mathrm{out}}| .
	\end{aligned}
	\label{rho out}
\end{equation}
Among them,  $|\psi_{(2,0),\mathrm{out}}\rangle,|\psi_{(1,1),\mathrm{out}}\rangle$ contribute to the vacuum output component, since two photons in modes $\hat{b}_3^{\text{ls}\dagger}, \hat{b}_4^{\text{ls}\dagger}$ are dissipated during the evolution. For similar reasons,  $|\psi_{(1,0),\mathrm{out}}\rangle,|\psi_{(0,1),\mathrm{out}}\rangle$ contribute to the single-photon output component, $|\psi_{(0,0),\mathrm{out}}\rangle$ contribute to the two-photon  output component. Since different $|\psi_{(i,j),\mathrm{out}}\rangle$ are statistically mixed, output density operator Eq. (\ref{rho out}) is in a block form, as shown in Table. \ref{table:1}. Through this example, the interconnection among photon loss, quantum coherence and decoherence is fully manifested. Within each DPLS, photon loss occurs while the coherence of quantum states is preserved. However, decoherence arises for the superposition of states belonging to different DPLSs. 

\section{Applications of DPLS Theory}
Once the theory of DPLS is established, we now consider its applications by several examples. In section \ref{re-explore}, some quantum interference phenomena in a common lossy system are revisited by the DPLS theory. The joint influences of quantum coherence, photon loss and quantum decoherence on input states are also investigated. In section \ref{generate W state}, the robust generation of W-state from various input states is demonstrated through constructing one-dimensional DPLSs in a lossy system. By tailoring DPLSs, quantum state evolution can be manipulated, thereby obtaining desired output states. These applications indicate that the DPLS theory can not only be applied to analyze quantum interference in lossy systems, but also serve as a guide to engineer non-Hermitian systems for quantum state preparation.

\subsection{Revisiting quantum interference phenomena by DPLS theory}\label{re-explore}
In this subsection, we apply DPLS theory to revisit quantum interference phenomena in a common class of lossy systems described by the following scattering matrix
\begin{equation}
	M=\frac{1}{2} \begin{pmatrix}
		1 & -1 \\
		-1 & 1 \\
	\end{pmatrix}.
	\label{25:25 lossy splitter}
\end{equation}
This kind of systems are often referred to as coherent perfect absorption (CPA) systems, which can be realized through lossy thin films \cite{25_roger2015coherent,26_roger2016coherent,27_altuzarra2017coherent} or coupled lossy waveguides \cite{46_zanotto2016design}. Applying SVD, we get 
\begin{equation}
	\begin{aligned}
		M & = U\Sigma V^{\dagger}\\
		& = \begin{pmatrix}
			-\frac{1}{\sqrt{2}} & \frac{1}{\sqrt{2}}\\
			\frac{1}{\sqrt{2}} & \frac{1}{\sqrt{2}} 
		\end{pmatrix}
		{\renewcommand{\arraystretch}{1.25}\setlength{\arraycolsep}{12pt} 
			\begin{pmatrix}
				1 & 0\\
				0 & 0
		\end{pmatrix}}
		\begin{pmatrix}
			-\frac{1}{\sqrt{2}} & \frac{1}{\sqrt{2}}\\
			\frac{1}{\sqrt{2}} & \frac{1}{\sqrt{2}} 
		\end{pmatrix}.
	\end{aligned}
\end{equation}
The two singular values are $\sigma_1=1$ and $\sigma_2=0$, thus a completely lossy input mode and a lossless input mode are defined as:
\begin{equation}
\begin{aligned}
	\hat{b}_1^{\mathrm{nls}} &= \frac{\hat{a}_1-\hat{a}_2}{\sqrt{2}}, \\
	\hat{b}_2^{\mathrm{ls}} &= \frac{\hat{a}_1+\hat{a}_2}{\sqrt{2}}.
\end{aligned}
\end{equation}
In the following, we consider quantum interference phenomena including the two-photon anti-Hong-Ou-Mandel (anti-HOM) interference \cite{13_vetlugin2022anti,16_ehrhardt2022observation,17_hong2024loss,18_vest2017anti,NatPhot_li2021}, the filtering of discrete variable (DV) states \cite{22_selim2025selective}, and the distillation of continuous variable (CV) states \cite{31_hardal2019quantum}. Since input states may be different in these examples, and the dimensions of DPLSs are determined by the number of input photons as well as lossless modes, only the relevant DPLSs are considered in each example.

\subsubsection{\textbf{Interference of discrete variable states}}
We first consider the Anti-HOM interference of photons \cite{13_vetlugin2022anti,16_ehrhardt2022observation,17_hong2024loss,18_vest2017anti,NatPhot_li2021} in this lossy system described by Eq. (\ref{25:25 lossy splitter}). In this case, when input is $\vert 11\rangle$, the output coincidence rate shows a peak, rather than the usual HOM-dip in lossless system. For the two-photon input state, there are three relevant DPLSs of the input Hilbert space:
\begin{equation}
	\begin{aligned}
		\mathcal{H}_{2}^{\mathrm{in}} &= \mathrm{Span}\left\{\frac{1}{\sqrt{2}}\left(\hat{b}_{2}^{\mathrm{ls}\dagger}\right)^{2}\,|0\rangle\right\},\\
		\mathcal{H}_{1}^{\mathrm{in}} &= \mathrm{Span}\left\{\frac{1}{\sqrt{2}}\hat{b}_{1}^{\mathrm{nls}\,\dagger}\hat{b}_{2}^{\mathrm{ls}\,\dagger}\,|0\rangle\right\},\\
		\mathcal{H}_{0}^{\mathrm{in}} &= \mathrm{Span}\left\{\frac{1}{\sqrt{2}}\left(\hat{b}_{1}^{\mathrm{nls}\,\dagger}\right)^{2}\,|0\rangle\right\}.
	\end{aligned}
\end{equation}
The input state $\vert 11\rangle$ is projected onto these DPLSs as
\begin{equation}
	\begin{aligned}
		|11\rangle
		&= \frac{|20\rangle+|02\rangle+\sqrt{2}\,|11\rangle-\left(|20\rangle+|02\rangle-\sqrt{2}\,|11\rangle\right)}{2\sqrt{2}}\\
		&= \frac{1}{\sqrt{2}}\left(|\psi_2\rangle-|\psi_0\rangle\right).
	\end{aligned}
	\label{11 projection}
\end{equation}
From Eq. (\ref{11 projection}), the probability of projecting the input state onto $\mathcal{H}_{2}$ and $\mathcal{H}_{0}$ are both $50\%$, and the two projections are 
\begin{equation*}
	\begin{aligned}
		|\psi_2\rangle &= \frac{|20\rangle+|02\rangle+\sqrt{2}\,|11\rangle}{2},\\
		|\psi_0\rangle &= \frac{|20\rangle+|02\rangle-\sqrt{2}\,|11\rangle}{2}.
	\end{aligned}
	\label{psi0 and psi2}
\end{equation*}
Considering the evolution of projections by Eq. (\ref{operator M}) as well as the simultaneous decoherence process, the output is thus
\begin{equation}
	\begin{aligned}
		\hat{\rho}_{\mathrm{out}}
		&= \frac{1}{2}\left(\hat{M}_{2}\,|\psi_{2}\rangle\langle \psi_{2}|\,\hat{M}_{2}^{\dagger}\right)
		+ \frac{1}{2}\left(\hat{M}_{0}\,|\psi_{0}\rangle\langle \psi_{0}|\,\hat{M}_{0}^{\dagger}\right)\\[4pt]
		&= \frac{1}{2}\,|00\rangle\langle 00|+ \frac{1}{2}\vert\psi_0\rangle\langle\psi_0\vert
	\end{aligned}
	\label{11 output}
\end{equation}
Since $\hat{M}_0 = \hat{S}_U \hat{S}_{V^{\dag}}=\bm{I}$, the output still contains $\vert\psi_0\rangle$, whose $\vert 11\rangle$ component leads to the non-zero outcome of $g^{(2)}$.

Using the above results, the anti-HOM peak can be further analyzed quantitatively, which is realized through comparing the output $g_{q,\mathrm{out}}^{(2)}$ of the indistinguishable two-photon input state with output $g_{c,\mathrm{out}}^{(2)}$ of the distinguishable two-photon input state \cite{NatPhot_li2021}, here $g^{(2)}
=
\langle \hat{a}_2^{\dagger}\hat{a}_1^{\dagger}\hat{a}_1\hat{a}_2 \rangle/
\langle \hat{a}_1^{\dagger}\hat{a}_1 \rangle
\langle \hat{a}_2^{\dagger}\hat{a}_2 \rangle $. It can be proved that in the two-photon input case, $\langle \hat{a}_2^\dagger \hat{a}_1^\dagger \hat{a}_1 \hat{a}_2 \rangle = P_{11}$.
For output Eq. (\ref{11 output}) of indistinguishable two-photon input, $g_{q,\mathrm{out}}^{(2)}$ can be calculated as
\begin{equation*}
	\begin{aligned}
		\left\langle \hat{a}_2^{\dagger}\hat{a}_1^{\dagger}\hat{a}_1\hat{a}_2 \right\rangle_{q}
		&= \operatorname{Tr}\!\left[\hat{\rho}_{\mathrm{out}}\, \hat{n}_1 \hat{n}_2\right]
		= \frac{1}{4},\\[4pt]
		\left\langle \hat{a}_1^{\dagger}\hat{a}_1 \right\rangle_{q}
		&= \left\langle \hat{a}_2^{\dagger}\hat{a}_2 \right\rangle_{q}
		= \frac{1}{2},\\[4pt]
		g^{(2)}_{q,\mathrm{out}}
		&= 1.
	\end{aligned}
\end{equation*}
In the distinguishable two-photon input case, each photon has a probability of $1/4$ to be transmitted or reflected, with the remaining $1/2$ probability to be absorbed. Therefore, the probability of obtaining output state $\vert11\rangle$ is
\begin{equation*}
	P_{|11\rangle,c} = P_{rr}+P_{tt}
	= 2\times\left(\frac{1}{4}\right)^2
	= \frac{1}{8}
\end{equation*}
Here, $P_{tt}$ and $P_{rr}$ represent probabilities that both photons are transmitted or reflected. Similarly, the probability of detecting 2,1 and 0 photons in mode $i (i=1,2)$ are $P(n_i=2) = \frac{1}{16}$, $P(n_i=1) = \frac{3}{8}$ and $P(n_i=0) = \frac{9}{16}$.
Therefore, $\langle \hat{a}_1^\dagger \hat{a}_1 \rangle_c
=
\langle \hat{a}_2^\dagger \hat{a}_2 \rangle_c
=
\frac{1}{2}, g_{c,\mathrm{out}}^{(2)}=\frac{1}{2}$, and the ratio $g_{q,\mathrm{out}}^{(2)}/g_{c,\mathrm{out}}^{(2)}=2$. While for both distinguishable and indistinguishable two-photon input states $\vert 11\rangle$, $g_{c,\mathrm{in}}^{(2)}=g_{q,\mathrm{in}}^{(2)}=1$, so 
\begin{equation*}
	\frac{g_{q,\mathrm{out}}^{(2)}}{g_{c,\mathrm{out}}^{(2)}} \;>\;
	\frac{g_{q,\mathrm{in}}^{(2)}}{g_{c,\mathrm{in}}^{(2)}},
\end{equation*}  
i.e., anti-HOM interference occurs in this lossy system, and the peak value $g_{q,\mathrm{out}}^{(2)}/g_{c,\mathrm{out}}^{(2)}=2$ equals to that in \cite{18_vest2017anti}.

Another frequently mentioned phenomenon in this system is the zero-or-two-photon absorption \cite{36_barnett1998quantum,39_tischler2018quantum}, where the input two photons of $\vert 11\rangle$ are either both absorbed, or non of them are absorbed, while exactly one photon absorption is impossible. Using DPLS, we can immediately find from Eq. (\ref{11 projection}) that $\vert11\rangle$ projects onto $\mathcal{H}_{2}^{\mathrm{in}}$ and $\mathcal{H}_{0}^{\mathrm{in}}$ both with $50\%$ probability. Therefore, during the evolution in each DPLS, either two photons are absorbed, or no photon is absorbed. Due to the quantum decoherence, the output is a mixture of zero-photon component and two-photon component, as shown by Eq. (\ref{11 output}). If input state is $(\vert20\rangle+\vert02\rangle)/\sqrt{2}$, zero-or-two-photon absorption will still take place, since it also projects onto $\mathcal{H}_{2}^{\mathrm{in}}$ and $\mathcal{H}_{0}^{\mathrm{in}}$ both with $50\%$ probability. Actually, if an input two-photon state only projects onto $\mathcal{H}_{2}^{\mathrm{in}}$ and $\mathcal{H}_{0}^{\mathrm{in}}$, the zero-or-two-photon absorption will always occur.

By further analyzing DPLSs of the system described by Eq. (\ref{25:25 lossy splitter}), we revisit the filtering of DV states \cite{22_selim2025selective}. When input photon number $N_p$ is fixed, all DPLSs are one-dimensional in the following form
\begin{equation}
	\mathcal{H}_{n}^{\mathrm{in}}=\mathrm{Span}\left\{
	\frac{1}{\sqrt{m!n!}}
	\left(\hat b_{1}^{\text{nls}\dagger}\right)^{m}
	\left(\hat b_{2}^{\text{ls}\dagger}\right)^{n}
	\left\vert0\right\rangle, m=N_p-n
	\right\}.
	\label{general form of DPLS}
\end{equation}
The projection in DPLS $\mathcal{H}_{n}^{\mathrm{in}}$ evolves according to Eq. (\ref{operator M}). Finally, it contributes to the $m$-photon output component
\begin{equation}
	\begin{aligned}
		\left\vert\psi_{n,\mathrm{out}}\right\rangle
		& =\frac{1}{\sqrt{m!}}\left(\hat{b}_{1}^{\text{nls}\dagger}\right)^{m}|0\rangle\\
		& =\sum_{k=0}^{m}(-1)^{k}\sqrt{\frac{1}{2^{m}}\binom{m}{k}}\;|m-k,k\rangle,
	\end{aligned}
	\label{m photon component}
\end{equation}
which is still a coherently superposed state. However, output components with different $m$ are only statistically mixed due to the occurrence of quantum decoherence. For different input states, the m-photon output components are all the same, only with different probabilities. Therefore, using the scattering matrix Eq. (\ref{25:25 lossy splitter}), states with the form of Eq. (\ref{m photon component}) can be filtered out, which are steady states in the m-photon output subspaces even under the influence of loss. These steady states have been obtained in the coupled-waveguide anti-PT system \cite{22_selim2025selective}. However, Hermitian interaction Hamiltonian is used to derive the steady state Eq. (\ref{m photon component}) there, which doesn't take the effect of loss into account. By contrast, the steady states can also be obtained using our DPLS theory, which fully considers the joint influence of quantum interference, photon loss and quantum decoherence.

By applying the theory of DPLS, we first physically re-explain the anti-HOM interference and two-photon absorption in the linear lossy system. Both of them occur because the input state $\vert11\rangle$ projects into DPLS $\mathcal{H}_{2}^{\mathrm{in}}$ and $\mathcal{H}_{0}^{\mathrm{in}}$. Then the filtering of DV states is also revisited, showing that projection in each DPLS evolves into a corresponding steady state. Therefore, by simply analyzing the DPLS structure of the system, quantum interference, photon number reduction and quantum decoherence can be investigated, thereby providing
a clear physical understanding of quantum state evolution in lossy systems.

\subsubsection{\textbf{Interference of continuous variable states}}
In the above examples, the theory of DPLS is applied to analyze quantum interference of discrete variable (DV) states in lossy systems. In fact, it can also be applied to solve the quantum interference of continuous variable (CV) states in lossy systems. Specifically, in this section we use DPLS to explain the mechanism of distillation of squeezed vacuum states \cite{31_hardal2019quantum}. Furthermore, it's shown that by engineering the DPLSs of the lossy system, in addition to the two identical input squeezed coherent states used in \cite{31_hardal2019quantum}, more types of input squeezed coherent states can be distilled to obtain squeezed vacuum states. 

First the original case is considered. The scattering matrix of the system is still Eq. (\ref{25:25 lossy splitter}), and the input is a direct product of two identical squeezed coherent states
\begin{equation*}
	\begin{aligned}
		|\psi_{\mathrm{in}}\rangle &= |\alpha,\xi\rangle \otimes |\alpha,\xi\rangle\\
		&=\hat S_{1}(\xi)\hat D_{1}(\alpha)\hat S_{2}(\xi)\hat D_2(\alpha)\,|0\rangle. 
	\end{aligned}
\end{equation*}
Squeezing operators and displacement operators are $\hat{S}_i(\xi)=\exp\left[-\frac{\xi}{2}\hat{a}_i^{\dagger2}+\text{H.c.}\right]$ and $\hat{D}_i(\alpha)=\exp\left(\alpha\hat{a}_i^{\dagger}-\text{H.c.}\right)$. The commutation relations satisfy $[\hat a_i,\hat a_j^\dagger]=[\hat a_i,\hat a_j]=\delta_{ij}$, so the input can be rewritten as 
\begin{equation*}
	\begin{aligned}
		|\psi_{\mathrm{in}}\rangle
		& = \hat S_{1}(\xi)\hat S_{2}(\xi)\hat D_{1}(\alpha)\hat D_{2}(\alpha)|0\rangle	\\
		& = 
		\exp\left[-\frac{\xi}{2}\left(\hat a_1^{\dagger 2}+\hat a_2^{\dagger 2}\right)+\mathrm{H.c.}\right]\cdot\\
		&\quad\exp\left[\alpha\left(\hat a_1^\dagger+\hat a_2^\dagger\right)-\mathrm{H.c.}\right]
		|0\rangle.
	\end{aligned}
\end{equation*}
The general form of DPLSs is the same as Eq. (\ref{general form of DPLS}), but now $m$ can be any non-negative integer, since $|\psi_{\mathrm{in}}\rangle$ is superposed by infinite number of Fock states. To project the input state onto DPLSs, we define $\hat{S}_i'(\xi)=\exp\left[-\frac{\xi}{2}\hat{b}_i^{\dagger 2}+\text{H.c.}\right]$ and $\hat{D}_i'(\alpha)=\exp\left(\alpha\hat{b}_i^{\dagger}-\text{H.c.}\right)$, the input state can be written as
\begin{equation*}
	\begin{aligned}
		|\psi_{\text{in}}\rangle
		&= \sum_{n} A_n \hat{S}_1'(\xi) \frac{\left(\hat{b}_{2}^{\text{ls}\dagger}\right)^n}{\sqrt{n!}} |0\rangle \\
		&= \sum_{n} A_n |\psi_n\rangle,
	\end{aligned}
\end{equation*}
where $\vert \psi_{n}\rangle$ is the projection in DPLS $\mathcal{H}_n^{\text{in}}$. The probability amplitudes $A_n$ satisfy
\begin{equation*}
	\begin{aligned}
		\sum_n A_n \frac{\left(\hat{b}_{2}^{\text{ls}\dagger}\right)^n}{\sqrt{n!}} |0\rangle &= \hat{S}_2'(\xi)\hat{D}_2'(\xi)|0\rangle, \\
		\quad \sum_n |A_n|^2 & = 1.
	\end{aligned}
\end{equation*}
The evolution in $\mathcal{H}_n^{\text{in}}$ is characterized by Eq. (\ref{operator M}), so the output of projection in $\mathcal{H}_n^{\text{in}}$ is 
\begin{equation}
	\begin{aligned}
		\hat{M}_n |\psi_n\rangle &= \hat{S}_U \hat{S}_{V^\dagger} \hat{S}_1'(\xi) |0\rangle \\
		&= \hat{S}_1'(\xi) |0\rangle,
	\end{aligned}
	\label{output in DPLS}
\end{equation}
where $UV^{\dagger}=\bm{I}$ is used in the above derivations. From Eq. (\ref{output in DPLS}), it's obvious that all projections in different DPLSs always evolve into the same squeezed vacuum state, therefore the final output is a pure squeezed vacuum state
\begin{equation*}
	\begin{aligned}
		\hat{\rho}_{\text{out}} &= \sum_n |A_n|^2 \hat{M}_n |\psi_n\rangle \langle\psi_n| \hat{M}_n^\dagger \\
		&= \hat{S}_1'(\xi) |0\rangle\langle 0| \left[\hat{S}_1'(\xi)\right]^{\dagger}.
	\end{aligned}
\end{equation*}
The coherent amplitude $\alpha$ in the input state is completely absorbed due to the photon dissipation process in each DPLS. Here, also by considering the state evolution in DPLS, the distillation of squeezed coherent states is analyzed without introducing ancilla modes. 

Actually, by engineering the DPLSs of the system, the distillation scheme can be generalized so that even squeezed coherent states with different coherent amplitudes can also be distilled. Consider the input state to be $|\psi_{\text{in}}\rangle = |\alpha_1,\xi\rangle \otimes |\alpha_2, \xi\rangle$, and SVD of a new scattering matrix is 
\begin{equation*}
	M = \begin{pmatrix} -\frac{1}{\sqrt{2}} & \frac{1}{\sqrt{2}} \\[1em] \frac{1}{\sqrt{2}} & \frac{1}{\sqrt{2}} \end{pmatrix}
	{\renewcommand{\arraystretch}{1.25}\setlength{\arraycolsep}{12pt}
		\begin{pmatrix} 1 & 0 \\ 0 & 0 \end{pmatrix}}
	\begin{pmatrix} -\cos\theta & \sin\theta \\ \sin\theta & \cos\theta \end{pmatrix}.
\end{equation*}
According to the above SVD, the new completely lossy input mode and lossless input mode are 
\begin{equation}
	\begin{aligned}
		\hat{b}_1^{\text{nls}} &= -\hat{a}_1 \cos\theta + \hat{a}_2 \sin\theta \\
		\hat{b}_2^{\text{ls}} &= \hat{a}_1 \sin\theta + \hat{a}_2 \cos\theta
	\end{aligned}
	\label{new b1 and b2}
\end{equation}
The DPLSs of the system still have the same form as in Eq. (\ref{general form of DPLS}), while $\hat{b}_1^{\text{nls}}$, $\hat{b}_2^{\text{ls}}$ are now in Eq. (\ref{new b1 and b2}) and $m\in\mathbb{N}$. So by modifying the scattering matrix, the DPLSs are engineered. Now the input state can be rewritten as 
\begin{equation}
	\begin{aligned}
		|\psi_{\text{in}}\rangle = & \hat{S}_1'(\xi)\hat{D}_1'(-\alpha_1\cos\theta+\alpha_2\sin\theta) \cdot\\
		& \hat{S}_2'(\xi)\hat{D}_2'(\alpha_1\cos\theta+\alpha_2\sin\theta) |0\rangle \\
		= & \sum_n C_n \vert\psi_n\rangle, 
	\end{aligned}
	\label{different squeezed coherent state}
\end{equation}
where projection $|\psi_n\rangle$ in $\mathcal{H}_n^{\text{in}}$ and the corresponding coefficient $C_n$ satisfy
\begin{equation*}
	\begin{aligned}
		|\psi_n\rangle & = \hat{S}_1'(\xi)\hat{D}_1'(-\alpha_1\cos\theta+\alpha_2\sin\theta) \frac{\left(\hat{b}_{2}^{\text{ls}\dagger}\right)^n}{\sqrt{n!}} |0\rangle,\\
		\sum_n C_n \frac{\left(\hat{b}_2^{\text{ls}\dagger}\right)^n}{\sqrt{n!}} |0\rangle 
		& = \hat{S}_2'(\xi)\hat{D}_2'(\alpha_1\cos\theta+\alpha_2\sin\theta) |0\rangle.\\
	\end{aligned}
\end{equation*}
Since photons in mode $\hat{b}_2^{\text{nls}\dagger}$ will be completely dissipated while those in mode $\hat{b}_1^{\text{ls}\dagger}$ will be preserved, if the desired output is still a squeezed vacuum state, then the coherent amplitude in $\hat{b}_1^{\text{ls}\dagger}$ should be nullified in Eq. (\ref{different squeezed coherent state}) to ensure that the output state has no coherent amplitude. Therefore, 
\begin{equation}
	\begin{aligned}
		-\alpha_1 \cos\theta + \alpha_2 \sin\theta = 0 \quad \rightarrow \quad \frac{\alpha_1}{\alpha_2} = \tan\theta
	\end{aligned}
\end{equation}
The final output is thus 
\begin{equation}
	\begin{aligned}
		\hat{\rho}_{\text{out}} & = \sum_n |C_n|^2 \, |\psi_{n,\text{out}}\rangle \langle \psi_{n,\text{out}}| \\
		& = \hat{S}_1'(\xi) |0\rangle \langle 0| \left[\hat{S}_1'(\xi)\right]^\dagger
	\end{aligned}
\end{equation}
which is still a pure squeezed vacuum state. Therefore, based on our theory, by modifying the scattering matrix to manipulate the DPLSs of the system, squeezed vacuum states can be distilled even from input squeezed coherent states with different coherent amplitudes. 

\subsection{One-dimensional DPLS for robust $W$-state preparation}\label{generate W state}
\begin{figure*}[t]
	\includegraphics[width=1\textwidth]{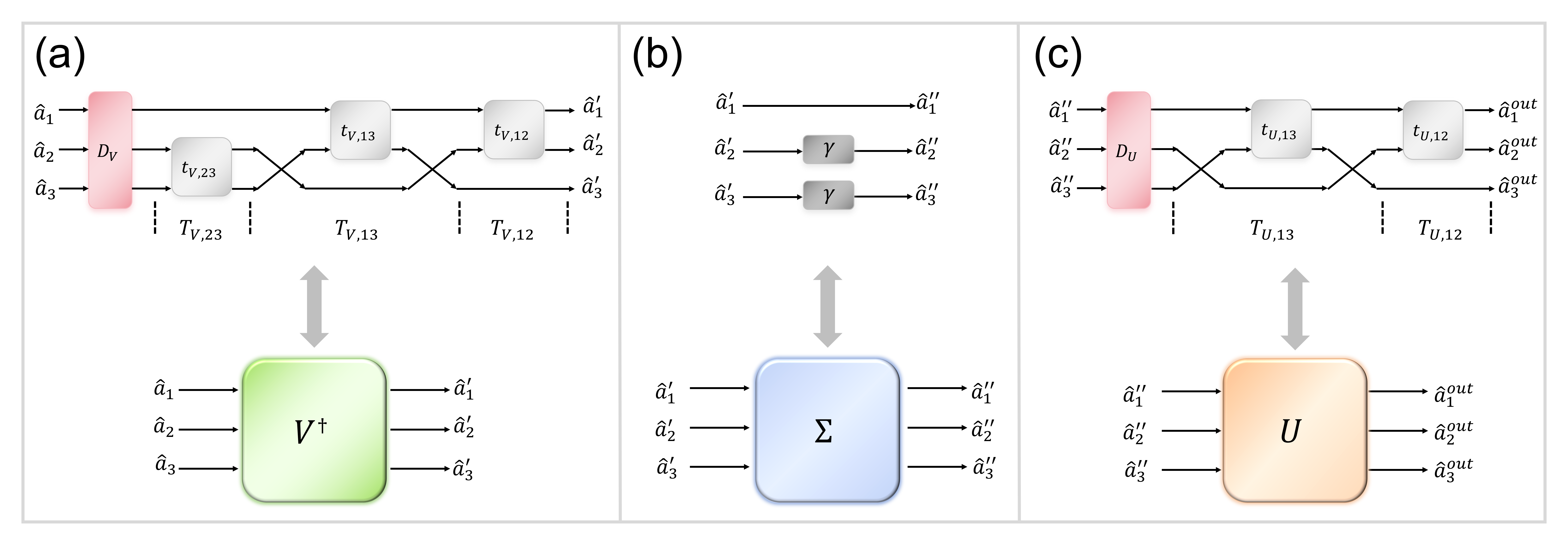}
	\caption{\label{fig:4} Constructed lossy system for robust W-state generation with three sequential parts: (a) Unitary transformation $V^{\dagger}$, (b) diagonal matrix $\Sigma$, and (c) unitary transformation $U$. 
	$V^{\dagger}$ and $U$ can be constructed by two-mode beam splitters $t$s and phase shifters $D$s. The two mode beam splitter $t$, together with an uncoupled mode, constitutes the corresponding three-mode unitary transformation $T$. $\Sigma$ can be constructed by adding adequate amount of loss $\gamma$ to modes $\hat{a}_{2}'$ and $\hat{a}_3'$. 
	}
	\label{system construction}
\end{figure*}
Besides being applied to analyze quantum interference in lossy systems, DPLS can also be used to direct the design of non-Hermitian systems. By tailoring DPLSs of the lossy systems, quantum coherence, photon loss and quantum decoherence can be controlled, thus target states can be prepared. In the following, taking the preparation of $W$-state based on DPLS theory as an example, the capability of using DPLS to design lossy systems for preparing specific quantum states is demonstrated. 

\begin{figure*}[t]
	\includegraphics[width=1\textwidth]{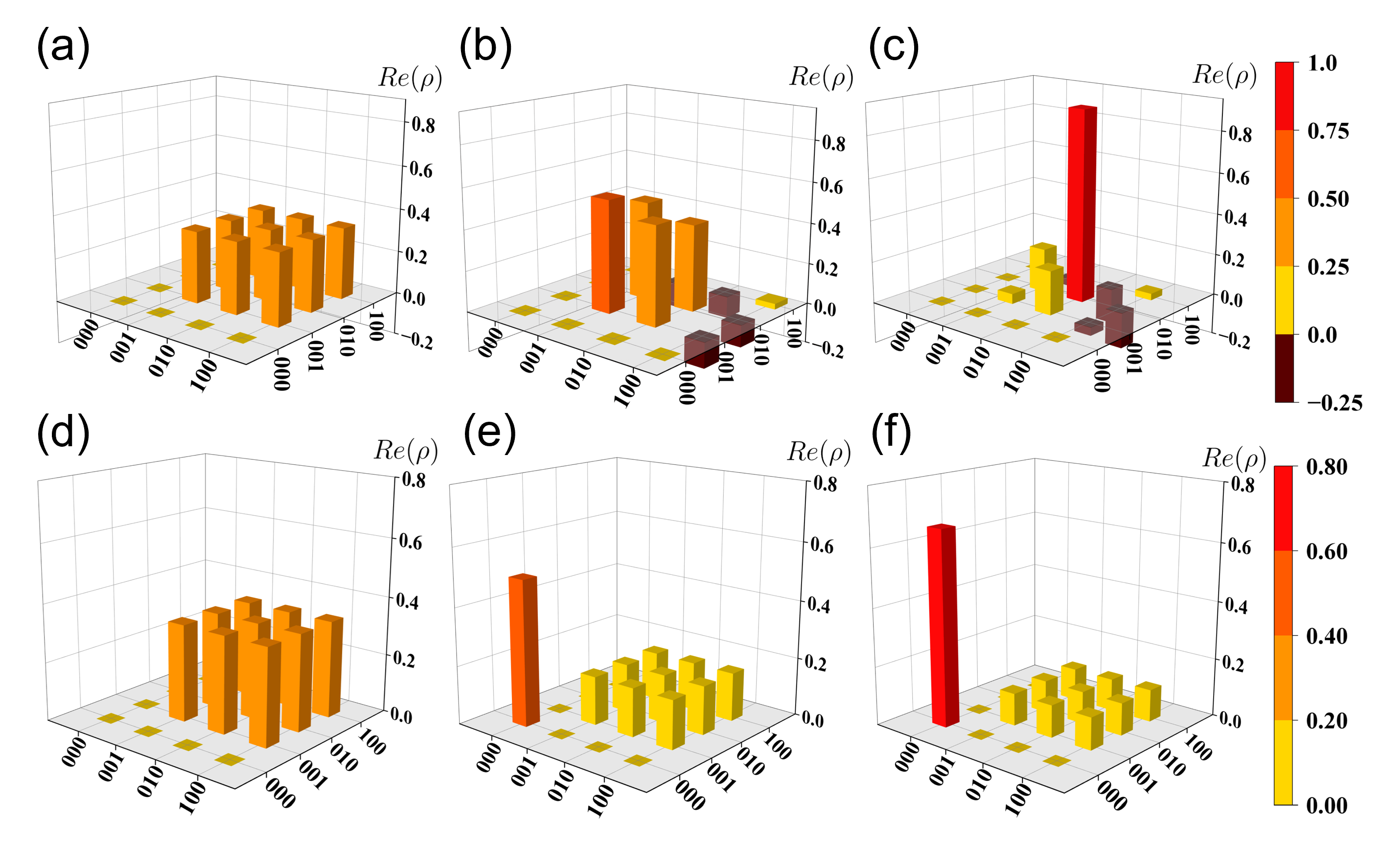}
	\caption{\label{fig:5} Preparation of W-state in lossless system and lossy system, x axis and y axis correspond to kets and bras respectively. The input states are (a), (d) $(\vert100\rangle-\vert010\rangle)/\sqrt{2}$, (b), (e) $\vert100\rangle$ and (c), (f) $(\sqrt{2}\vert100\rangle+\vert001\rangle)/\sqrt{3}$. (a)-(c) are results for lossless system, where W-state can no longer be obtained once the input varies. (d)-(f) are results for lossy system, by engineering DPLSs of the system, W-state can be prepared for various input states. } 
	\label{W result}
\end{figure*}

$\vert W\rangle$ is a multipartite entangled state with the general form
\begin{equation*}
	\begin{aligned}
		|W_N\rangle = \frac{1}{\sqrt{N}} \left( |10\ldots0\rangle + |010\ldots0\rangle + \cdots + |0\ldots01\rangle \right)
	\end{aligned}.
\end{equation*}
Its entanglement is robust against loss \cite{40_guhne2009entanglement}. 
Previously, $W$-state is mainly generated in lossless systems \cite{2023W_Nano_Lett,2009W_Science,2014W_Nat_Photon}, where specific input is required. For example, consider transferring a bipartite entangled state $\vert\psi_-\rangle=(\vert100\rangle-\vert 010\rangle)/\sqrt{2}$ into a tripartite entangled $W$-state $\vert W\rangle=(\vert100\rangle+\vert010\rangle+\vert001\rangle)/\sqrt{3}$ by the following unitary beam splitter
\renewcommand{\arraystretch}{1.8}  
\begin{equation*}
	T = \begin{pmatrix}
		\frac{1}{6} & \frac{1}{6} + \sqrt{\frac{2}{3}} & -\frac{1}{3} + \frac{1}{\sqrt{6}} \\
		\frac{1}{6} - \sqrt{\frac{2}{3}} & \frac{1}{6} & -\frac{1}{3} - \frac{1}{\sqrt{6}} \\
		-\frac{1}{3} - \frac{1}{\sqrt{6}} & -\frac{1}{3} + \frac{1}{\sqrt{6}} & \frac{2}{3}
	\end{pmatrix}.
\end{equation*}
The success probability is $100\%$, as shown in Fig. \ref{W result} (a). However, only specific input state can be converted to the desired output $W$-state by this unitary transformation. Once the input changes, the output is no longer a $W$-state. For instance, if input states are $\vert 100\rangle$ and $(\sqrt{2}\vert100\rangle+\vert001\rangle)/\sqrt{3}$, the output single-photon states are not $W$-states any more, shown in Fig. \ref{W result} (b) and Fig. \ref{W result} (c).

Nevertheless, $\vert \psi_-\rangle$ can also be transferred into $\left\vert W\right\rangle$ with $100\%$ success probability by a lossy system, as long as $\vert \psi_-\rangle$ belongs to DPLS $\mathcal{H}_{\vec{0}}^{\text{in}}$ of the system. We now consider 
\begin{equation}
	\begin{aligned}
		M = U\Sigma V^{\dagger} 
		&= \left(\begin{array}{c@{\hspace{1cm}}c@{\hspace{1cm}}c}
			-\dfrac{1}{\sqrt{6}} & \dfrac{1}{\sqrt{6}} & 0 \\[2mm]
			-\dfrac{1}{\sqrt{6}} & \dfrac{1}{\sqrt{6}} & 0 \\[2mm]
			-\dfrac{1}{\sqrt{6}} & \dfrac{1}{\sqrt{6}} & 0
		\end{array}\right),
	\end{aligned}
	\label{lossy system for W generation}
\end{equation}
where $U, \Sigma, V^{\dagger}$ are
\renewcommand{\arraystretch}{1.5}
\setlength{\arraycolsep}{10pt}
\begin{align}
	U & = \begin{pmatrix}
		-\dfrac{1}{\sqrt{3}} & \dfrac{1}{\sqrt{2}} & \dfrac{1}{\sqrt{6}} \\[3mm]
		-\dfrac{1}{\sqrt{3}} & -\dfrac{1}{\sqrt{2}} & \dfrac{1}{\sqrt{6}} \\[3mm]
		-\dfrac{1}{\sqrt{3}} & 0 & -\dfrac{\sqrt{2}}{\sqrt{3}}
	\end{pmatrix},\notag
\end{align}
\begin{align}
	V^{\dagger} & = \begin{pmatrix}
		\dfrac{1}{\sqrt{2}} & -\dfrac{1}{\sqrt{2}} & 0 \\[3mm]
		\dfrac{1}{\sqrt{3}} & \dfrac{1}{\sqrt{3}} & \dfrac{1}{\sqrt{3}} \\[3mm]
		\dfrac{1}{\sqrt{6}} & \dfrac{1}{\sqrt{6}} & -\dfrac{\sqrt{2}}{\sqrt{3}}
	\end{pmatrix},\notag
\end{align}
\begin{align}
	\setlength{\arraycolsep}{16pt}
	\Sigma & = \begin{pmatrix}
		\hspace{0.6cm} 1 & \hspace{0.3cm} 0 \hspace{0.3cm} & 0 \hspace{0.6cm} \\[2mm]
		\hspace{0.6cm} 0 & \hspace{0.3cm} 0 \hspace{0.3cm} & 0 \hspace{0.6cm} \\[2mm]
		\hspace{0.6cm} 0 & \hspace{0.3cm} 0 \hspace{0.3cm} & 0 \hspace{0.6cm}
	\end{pmatrix}.\notag
\end{align}

According to the matrix decomposition in \cite{41_reck1994experimental}, $V^{\dagger}$ and $U$ can be constructed using two-port beam splitters $t$ and phase shifters $D$ as
\begin{equation}
	\begin{aligned}
		V^\dagger & = T_{V,12} \, T_{V,13} \, T_{V,23} \, D_V\\
		U & = T_{U,12} \, T_{U,13} \, D_U
	\end{aligned}
	\label{UVSigma for W generation}
\end{equation}
Here, $T_{V,ij}$ and $T_{U,ij}$ are actually three-mode lossless beam splitters composed by the corresponding two-mode beam splitters $t$ and an uncoupled lossless mode, as shown in Fig. (\ref{system construction}) (a) and (c). Their detailed forms are in Appendix \ref{T&D}.
While for the diagonal matrix $\Sigma$, it can be realized by introducing adequate amount of loss $\gamma$ to modes $\hat{a}_{2}'$ and $\hat{a}_3'$, so that photons initially in these two modes are completely dissipated during the evolution.These three matrices can be realized as Fig. \ref{system construction} (a)-(c). 

According to SVD Eq. (\ref{lossy system for W generation}), there is only one lossless input mode $\hat{b}_1^{\text{nls}}$ and two completely lossy input modes $\hat{b}_{2,3}^{\text{ls}}$
\begin{equation}
\begin{aligned}
	\hat{b}_1^{\text{nls}\dagger} &= \frac{\hat{a}_1^\dagger - \hat{a}_2^\dagger}{\sqrt{2}}, \\[6pt]
	\hat{b}_2^{\text{ls}\dagger} &= \frac{\hat{a}_1^\dagger + \hat{a}_2^\dagger + \hat{a}_3^\dagger}{\sqrt{3}}, \\[6pt]
	\hat{b}_3^{\text{ls}\dagger} &= \frac{\hat{a}_1^\dagger + \hat{a}_2^\dagger - 2\hat{a}_3^\dagger}{\sqrt{6}}. 
\end{aligned}
\end{equation}
When the number of input photons $N_p$ is fixed, all DPLSs are one-dimensional according to Eq. (\ref{DPLS dimension}), and can be written as 
\begin{equation}
	\begin{aligned}
		H_{\vec{n}}^{\text{in}} = \text{Span}\biggl\{ & 
		\frac{1}{\sqrt{m_1! \, n_2! \, n_3!}} 
		\left( \hat{b}_1^{\text{nls}\dagger} \right)^{m_1}
		\left( \hat{b}_2^{\text{ls}\dagger} \right)^{n_2}
		\left( \hat{b}_3^{\text{ls}\dagger} \right)^{n_3} |0\rangle,\\
		& m_1 = N_p - (n_2 + n_3) \biggr\}.
	\end{aligned}
\end{equation}
Obviously, $\vert\psi_-\rangle=(\vert100\rangle-\vert 010\rangle)/\sqrt{2}$ belongs to the lossless subspace $\mathcal{H}_{(0,0)}^{\text{in}}$ of the system , and the corresponding evolution operator is $\hat{M}_{(0,0)}=\hat{S}_U\hat{S}_{V^{\dagger}}$. Since for this system, $UV^{\dagger}=T$, $\vert\psi_-\rangle$ will be unitarily transformed into $\vert W\rangle$ with $100\%$ success probability, as shown in Fig. \ref{W result} (d). 

More generally, for an arbitrary input state $\vert \psi_{\text{in}}\rangle$ with $N_p$ photons, consider its projections $\vert\psi_{\vec{n}}\rangle$ in DPLSs with $n_2+n_3=N_p-1$, i.e., DPLSs where only one photon is in the lossless input mode,
\begin{equation}
	|\psi_{\vec{n}}\rangle = \frac{1}{\sqrt{n_2! \, n_3!}} \hat{b}_1^{\text{nls}\dagger} \left( \hat{b}_2^{\text{ls}\dagger} \right)^{n_2} \left( \hat{b}_3^{\text{ls}\dagger} \right)^{n_3} |0\rangle.
\end{equation}
These projections contribute to the single-photon components of the output state, which can be obtained using Eq. (\ref{operator M}) as 
\begin{equation}
	\begin{aligned}
		\hat{M}_{\vec{n}} |\psi_{\vec{n}}\rangle = \hat{S}_U \hat{S}_{V^\dagger} \hat{b}_1^{\text{nls}\dagger} |0\rangle =  |W\rangle.
	\end{aligned}
\end{equation}
As a result, the single-photon component of the output state always takes the form of a $W$-state. This is due to the fact for a lossy system containing only one lossless mode and all other modes subject to complete loss, when the number of input photons is $N_p$, all relevant DPLSs are one-dimensional according to Eq. (\ref{DPLS dimension}), meaning that different input states have the same projections in DPLSs with $n_2+n_3=N_p-1$. During the evolution, photons distributed in modes $\hat{b}_{2,3}^{\text{ls}\dagger}$ are deterministically dissipated, while the single photon in $\hat{b}_1^{\text{nls}\dagger}$ evolves into $\vert W\rangle$, contributing to the single-photon component of the output state. Output components with different photon number are statistically mixed, since they come from projections in different DPLSs. Then, by post-selecting the single-photon output component, $W$-state can be obtained. For example, when input states are $(\vert100\rangle-\vert010\rangle)/\sqrt{2}$, $\vert 100\rangle$ and $(\sqrt{2}\vert100\rangle+\vert 001\rangle)/\sqrt{3}$, $\vert W\rangle$ can always be post-selected with success probabilities $100\%$, $50\%$, and $33.3\%$ respectively, as shown in Fig. \ref{W result} (d)-(f). Therefore, by tailoring DPLS, controlled quantum coherence, photon loss as well as quantum decoherence can be realized to robustly prepare target states in lossy systems.

The applications in this section show that our DPLS theory can not only be applied to reveal the influences of quantum decoherence and photon number reduction on quantum interference, but also serve as a guide to engineer non-Hermitian systems for robust quantum state preparation. Although we mainly consider the two-mode and three-mode systems in these examples, our theory can be naturally applied to higher dimensional lossy systems, thus showing potentials in high-dimensional entanglement preparation and multi-qubit quantum gate implementation in non-Hermitian systems.

\section{conclusion} 
In summary, we have proposed the general theory of DPLS for quantum interference in lossy systems. By performing SVD of the scattering matrix with singular values only in $\{0,1\}$, completely lossy and lossless input modes of the system are defined, thereby decomposing the input Hilbert space into a direct sum of DPLSs according to the number of photons in the completely lossy input modes.  Quantum state evolution is then analyzed by evolving the projected state within each DPLS separately and subsequently constructing a statistical mixture of the resulting states. Based on our theory, quantum interference phenomena including the anti-HOM interference and the distillation of quantum states are revisited. Moreover, by constructing one-dimensional DPLSs to precisely control the quantum interference, robust $W$-state generation from various input states is demonstrated in a lossy system. Our theory clearly elucidates the interplay among quantum interference, photon number reduction and quantum decoherence by investigating the loss-induced subspaces of the system: state within each DPLS evolves coherently even under the influence of photon loss, while output state is a statistical mixture of all evolved states in different DPLS. It also provides a promising framework for quantum state preparation, quantum logic operations, and other quantum information processing tasks in lossy systems.

The data that support the findings of this study are available from the corresponding author upon resonable request.

\section*{Acknowledgments}
This work is supported by the National Natural Science Foundation of China under Grant No. 12474370 and No. U25D9003 and the Quantum Science and Technology-National Science and Technology Major Project No. 2021ZD0301500.

\appendix
\section{Operation of diagonal matrix $\Sigma$ on quantum states}\label{Appendix Sigma}
The key step in the DPLS theory is to elucidate how losses influence quantum state evolution. According to SVD, the non-unitary diagonal matrix $\Sigma$ fully characterizes the losses in the system, therefore in this section we derive the output state $\hat{\rho}_{\text{mid2}}$ obtained by applying the transformation $\Sigma$ to the state $\vert\psi_{\text{mid1}}\rangle$. It will be shown that after the operation of $\Sigma$, quantum coherence is completely preserved for components of $\vert\psi_{\text{mid1}}\rangle$ with the same photon number distribution $\vec{n}$ in lossy modes, while those with different $\vec{n}$ become statistical mixtures.

According to Eq. (\ref{S operates on the input state}) and Eq. (\ref{psi mid1}), $|\psi_{\text{mid1}}\rangle$ can be written in the density operator form as  
\begin{equation}
	\begin{aligned}
		\hat{\rho}_{\text{mid1}} &= |\psi_{\text{mid1}} \rangle \langle \psi_{\text{mid1}} | \\
		&= \sum_{\vec{n}} \sum_{\vec{n}'} \sum_{\vec{m}} \sum_{\vec{m}'} D_{\vec{m}\vec{n}} D_{\vec{m}'\vec{n}'}^* \; |m_1 \ldots n_N \rangle \langle m_1' \ldots n_N' |, 
	\end{aligned}
	\label{rho mid1}
\end{equation}
where $D_{\vec{m}\vec{n}}$ is the probability amplitude for projecting $|\psi_{\text{mid1}} \rangle$ onto basis state $|m_1 \ldots m_k n_{k+1}\ldots n_N \rangle$. Then, 
By introducing $N$ input and $N$ output bosonic ancilla modes $\hat{a}_{Ai}'$ and $\hat{a}_{Ai}''$ initially in the vacuum states to describe the loss, the input-output relation of the extended system is \cite{38_hernandez2022generalized}
\begin{equation}
	\begin{pmatrix}
		\vec{\hat{a}}''\\ \vec{\hat{a}}_{A}''
	\end{pmatrix}
	=
	\begin{pmatrix}
		\Sigma & \Sigma_A \\ -\Sigma_A & \Sigma
	\end{pmatrix}
	\begin{pmatrix}
		\vec{\hat{a}}'\\ \vec{\hat{a}}_{A}'
	\end{pmatrix},
\end{equation}
where $\vec{\hat{a}}'=(\hat{a}_1',\ldots,\hat{a}_N')$, and $\vec{\hat{a}}_A'=(\hat{a}_{A1}',\ldots,\hat{a}_{AN}')$. Definitions are similar for $\vec{\hat{a}}''$ and $\vec{\hat{a}}_A''$. Matrix $\Sigma_A = \operatorname{diag}\left( 0_1, \ldots, 0_k, 1_{k+1}, \ldots, 1_N \right)=\sqrt{\bm{I}-\Sigma^2}$. Accordingly,  
\begin{equation}
	\hat{a}_i' \to 
	\begin{cases}
		\hat{a}_i'', & 1 \leq i \leq k \\[4pt]
		-\hat{a}_{Ai}'', & k+1 \leq i \leq N
	\end{cases}.
\end{equation}
Therefore, $\hat{\rho}_{\text{mid1}}$ is transformed into $\hat{\rho}_{\text{mid2}}^{\text{full}}$ with ancilla modes as 
\begin{equation}
	\begin{aligned}
		\hat{\rho}_{\text{mid}2}^{\text{full}} =& \sum_{\vec{n}} \sum_{\vec{n}'} \sum_{\vec{m}} \sum_{\vec{m}'}  (-1)^{\sum_{i=k+1}^{N} (n_i + n_i')} D_{\vec{m}\vec{n}} D_{\vec{m}'\vec{n}'}^*\cdot \\
		&  |m_1 \dots m_k 0 \dots 0\rangle \langle m_1' \dots m_k' 0 \dots 0|  \otimes\\
		& |0 \dots 0 n_{k+1} \dots n_N \rangle_A \langle 0 \dots 0 n_{k+1}' \dots n_N' |.
	\end{aligned}
	\label{rho2full}
\end{equation}
Here subscript $A$ represents ancilla modes. From Eq. (\ref{rho2full}), photons in modes $\hat{a}_i'^{\dagger} (k+1\le i \le N)$ are all transformed into ancilla modes, which can't be observed in the output optical field. To obtain the output optical state $\hat{\rho}_{\text{mid2}}$, ancilla modes should be traced out so that
\begin{widetext}
	\begin{equation}
		\begin{aligned}
			\hat{\rho}_{\text{mid}2} &= \text{Tr}_A\left(\hat{\rho}_{\text{mid}2}^{\text{full}}\right)= \sum_{(q_1 \dots q_N)} \left(I_N \otimes {}_A \langle 0 \dots 0 q_{k+1} \dots q_N | \right) \hat{\rho}_{\text{mid}2}^{\text{full}}\left(I_N \otimes |0 \dots q_{k+1} \dots q_N \rangle_A \right) \\
			&= \sum_{\vec{q}} \sum_{\vec{n}} \sum_{\vec{n}'} \sum_{\vec{m}} \sum_{\vec{m}'} (-1)^{\sum_{i=k+1}^{N} (n_i + n_i')} D_{\vec{m}\vec{n}} D_{\vec{m}'\vec{n}'}^*|m_1 \dots m_k 0 \dots 0\rangle \langle m_1' \dots m_k' 0 \dots 0| \prod_{l=k+1}^{N} \delta_{q_l n_l} \delta_{q_l n_l'} \\
			&= \sum_{\vec{n}} \sum_{\vec{m}} \sum_{\vec{m}'} D_{\vec{m}\vec{n}} D_{\vec{m}'\vec{n}}^*|m_1 \dots m_k 0 \dots 0\rangle\langle m_1' \dots m_k' 0 \dots 0| \\
			&= \sum_{\vec{n}} P_{\vec{n}} \left[ \sum_{\vec{m}} \sum_{\vec{m}'} \left( \frac{D_{\vec{m}\vec{n}} D_{\vec{m}'\vec{n}}^*}{P_{\vec{n}}} \right) |m_1 \dots m_k 0 \dots 0\rangle \langle m_1' \dots m_k' 0 \dots 0| \right] \\
			&= \sum_{\vec{n}} P_{\vec{n}} \left( \sum_{\vec{m}} \frac{D_{\vec{m}\vec{n}}}{\sqrt{P_{\vec{n}}}} |m_1 \dots m_k 0 \dots 0\rangle \right) \left( \sum_{\vec{m}'} \frac{D_{\vec{m}'\vec{n}}^*}{\sqrt{P_{\vec{n}}}} \langle m_1' \dots m_k' 0 \dots 0| \right) \\
			&= \sum_{\vec{n}} P_{\vec{n}} |\psi_{\vec{n},\text{mid}2} \rangle \langle \psi_{\vec{n},\text{mid}2} |.
		\end{aligned}
		\label{rho mid2}
	\end{equation}
\end{widetext}

In the above equation, $P_{\vec{n}}=\sum_{\vec{m}}\vert D_{\vec{m}\vec{n}}\vert^2$ equals to the probability of projecting the input state onto DPLS $\mathcal{H}_{\vec{n}}^{\text{in}}$. Consequently, all photons in modes $\hat{a}_i'^{\dagger} (k+1\le i \le N)$ are dissipated, while those in modes $\hat{a}_i'^{\dagger} (1\le i \le k)$ are conserved. Mapping the operators $\hat{a}_i'^{\dagger}$ to the input of the system by unitary matrix $V$, $\hat{a}_i'^{\dagger} (k+1\le i \le N)$ are transformed into completely lossy input modes $\hat{b}_i^{\text{ls}\dagger}(k+1\le i\le N)$, and $\hat{a}_i'^{\dagger} (1\le i \le k)$ are transformed into lossless input modes $\hat{b}_i^{\text{nls}\dagger}(1\le i\le k)$. Therefore, photons initially in modes $\hat{b}_i^{\text{ls}\dagger}$ will be completely dissipated, while photons initially in modes $\hat{b}_i^{\text{nls}\dagger}$ will be completely transmitted. Moreover, from Eq. (\ref{rho mid1}) and Eq. (\ref{rho mid2}), quantum coherence is preserved among components of  $\hat{\rho}_{\text{mid1}}$ with the same photon number distribution $\vec{n}$. These components are then evolved into 
\begin{equation}
	\begin{aligned}
		\left[ \sum_{\vec{m}} \frac{D_{\vec{m}\vec{n}}}{\sqrt{P_{\vec{n}}}} |m_1 \dots m_k 0 \dots 0\rangle \right]
		\left[ \sum_{\vec{m}'} \frac{D_{\vec{m}'\vec{n}}^*}{\sqrt{P_{\vec{n}}}} \langle m_1' \dots m_k' 0 \dots 0| \right],
	\end{aligned}
\end{equation}
whereas components with different photon number distributions $\vec{n}$ become a statistical mixture, with their corresponding probabilities being $P_{\vec{n}}$, and $\sum_{\vec{n}} P_{\vec{n}}= 1$.

\section{Proof that all DPLSs form a direct sum decomposition of the input Hilbert space}\label{Direct sum decomposition of Hilbert space}
In the section, we prove that all DPLSs $\mathcal{H}_{\vec{n}}^{\text{in}}$ form a direct sum decomposition of the input Hilbert space
\begin{equation}
	\mathcal{H}^{\text{in}} = \mathcal{H}_{\vec{n}_1}^{\text{in}} \oplus \mathcal{H}_{\vec{n}_2}^{\text{in}} \oplus \dots ,
\end{equation}
which insures that any input state can be uniquely superposed by states in each DPLS. This property can be proved through verifying the following two conditions \cite{45_hoffmann1971linear}:
\begin{enumerate}
	\item All DPLSs $\mathcal{H}_{\vec{n}_i}^{\text{in}}$ span the whole Hilbert space $\mathcal{H}^{\text{in}}$:
	\begin{equation}
		\mathcal{H}^{\text{in}} = \mathcal{H}_{\vec{n}_1}^{\text{in}} + \mathcal{H}_{\vec{n}_2}^{\text{in}} + \dots.
	\end{equation}
	\item $\mathcal{H}_{\vec{n}_i}^{in}$ are independent to each other:
	\begin{equation}
		0 = \sum_i |\psi_{\vec{n}_i} \rangle, \quad |\psi_{\vec{n}_i} \rangle \in H_{\vec{n}_i}^{\text{in}} \iff \forall |\psi_{\vec{n}_i} \rangle=0.
	\end{equation}
\end{enumerate}

For condition 1, according to Eq. (\ref{basis state}), basis states of
DPLS $\mathcal{H}_{\vec{n}}^{in}$ is actually transformed from Fock state $\vert m_1\cdots m_k n_{k+1}\cdots n_N\rangle$ by unitary matrix $V$. Therefore,  The input Hilbert space $\mathcal{H}^{\text{in}}$ can also be spanned by these basis states as  
\begin{equation}
	\mathcal{H}^{\text{in}} = \text{Span}\left\{\vert\varphi_{\vec{m}\vec{n}}\rangle,\forall\vec{m},\vec{n}\right\}.
\end{equation}
Consequently, arbitrary input state $\vert \psi_{\text{in}}\rangle \in \mathcal{H}^{\text{in}}$ can be written as 
\begin{equation}
	\begin{aligned}
		|\psi_{\text{in}} \rangle &= \sum_{\vec{m}} \sum_{\vec{n}} D_{\vec{m}\vec{n}} |\varphi_{\vec{m}\vec{n}} \rangle \\
		&= \sum_{\vec{n}} \left( \sum_{\vec{m}} D_{\vec{m}\vec{n}} |\varphi_{\vec{m}\vec{n}} \rangle \right),
	\end{aligned}
	\label{psi_in superposed by states in DPLS}
\end{equation}
where $\sum_{\vec{m}} D_{\vec{m}\vec{n}} |\varphi_{\vec{m}\vec{n}} \rangle$ is an unnormalized state in DPLS $\mathcal{H}_{\vec{n}}^{\text{in}}$. According to Eq. (\ref{psi_in superposed by states in DPLS}), arbitrary input state can be superposed by quantum states from different DPLS, therefore, 
\begin{equation}
	\mathcal{H}^{\text{in}} = \sum_{\vec{n}_i} \mathcal{H}_{\vec{n}_i}^{\text{in}},
\end{equation}
condition 1 is proved.

Condition 2 can be verified by contradiction. Assume that there are non-zero states $\vert\psi_{\vec{n}_i}\rangle$ in $\mathcal{H}_{\vec{n}_i}^{\text{in}}$ that satisfy:
\begin{equation}
	|\psi_{\vec{n}_1} \rangle + |\psi_{\vec{n}_2} \rangle + \cdots + |\psi_{\vec{n}_N} \rangle = 0.
\end{equation}
After transposing $|\psi_{\vec{n}_N} \rangle$ to the other side and multiplying by $\langle\psi_{\vec{n}_N}\vert$, 
\begin{equation}
	\langle \psi_{\vec{n}_N} | \psi_{\vec{n}_1} \rangle + \langle \psi_{\vec{n}_N} | \psi_{\vec{n}_2} \rangle + \cdots + \langle \psi_{\vec{n}_N} | \psi_{\vec{n}_{N-1}} \rangle = - \langle \psi_{\vec{n}_N} | \psi_{\vec{n}_N} \rangle.
\end{equation}
According to Eq. (\ref{basis state}), all basis states are orthogonal to each other, so DPLSs spanned by different sets of $\left\{\vert\varphi_{\vec{m}\vec{n}}\rangle\right\}$ are also orthogonal to each other.
Therefore, $\langle \psi_{\vec{n}_N} | \psi_{\vec{n}_N} \rangle=0$, which implies that $| \psi_{\vec{n}_N} \rangle=0$. Repeating the above steps, we obtain
\begin{equation}
	|\psi_{\vec{n}_1} \rangle = |\psi_{\vec{n}_2} \rangle = \cdots = |\psi_{\vec{n}_N} \rangle = 0,
\end{equation}
leading to a contradiction with our assumption. Therefore, condition 2 is verified, thus the statement that all DPLSs form a direct sum decomposition of $\mathcal{H}^{\text{in}}$ is proved.

\section{Quantum interference described by scattering matrix with singular values $\sigma\in [0,1]$}\label{SV between 0 & 1}
In the main text, we considered non-unitary scattering matrices with singular values $\sigma\in \{0,1\}$. The characteristic of such a system is that the input Hilbert space can be decomposed into several DPLSs. In each DPLS, quantum states will deterministically dissipate  photons in certain modes during evolution, while quantum coherence is maintained for these states even in the presence of loss. For the superposition of states in different DPLSs, the output is a statistical mixture. Here, the DPLS theory is extended to a more general case,  where scattering matrices of lossy linear systems have singular values $\sigma\in \left[0,1\right]$. It's shown that by introducing ancilla bosonic modes, the enlarged Hilbert space $\mathcal{H}^{\text{in}}$ can also be decomposed into several DPLSs for analyzing quantum interference.

Suppose the SVD of an $N$-dimensional scattering matrix gives $M=U\Sigma V^{\dagger}$. The diagonal matrix $\Sigma$ is now $\text{diag}\left(1,\cdots,1,\sigma_1,\cdots,\sigma_j,0,\cdots,0\right)$, where the first $k$ diagonal elements are 1, the last $(N-j-k)$ diagonal elements are 0. The remaining $j$ elements $\sigma_1,\cdots,\sigma_j\in (0,1)$ indicate that photons in input modes $\hat{a}_{k+i} (i=1,\cdots,j)$ will be probabilistically dissipated. Our DPLS theory is also capable of analyzing quantum interference in such a lossy system, as long as the scattering matrix is embedded in a larger matrix with singular values $\sigma \in \{0,1\}$. This can be achieved by introducing ancilla modes $\hat{g}_i$ and $\hat{g}_i^{\text{out}}$ for each input mode $\hat{a}_{k+i}$ and output mode $\hat{a}_{k+i}^{\text{out}} $ corresponding to $\sigma_i\in (0,1)(i=1,\cdots,j)$, and then constructing two-mode non-unitary transformations $S_i$ satisfying
\begin{equation}
	S_{i} = {\renewcommand{\arraystretch}{1}\setlength{\arraycolsep}{8pt}\begin{pmatrix} \sigma_i & \sqrt{1 - \sigma_i} \\ 0 & 0 \end{pmatrix}}.
\end{equation}
between them. It can be proved that the non-unitary matrix $S_i$ has singular values $\sigma\in \{0,1\}$. Then, by arranging the input and output with ancilla modes as
\begin{equation}
	\begin{aligned}
		\vec{\hat{a}}^{\text{in}} &= \left(\hat{a}_1, \dots, \hat{a}_N, \hat{g}_1, \dots, \hat{g}_j\right)^T ,\\
		\vec{\hat{a}}^{\text{out}} &= \left(\hat{a}_1^{\text{out}}, \dots, \hat{a}_N^{\text{out}}, \hat{g}_1^{\text{out}}, \dots, \hat{g}_j^{\text{out}}\right)^T ,
	\end{aligned}
\end{equation}
$U,\Sigma, V^{\dagger}$ are accordingly enlarged as

\begin{equation}
	\begin{aligned}
		U &\to U_{\text{ext}} = {\renewcommand{\arraystretch}{1}\setlength{\arraycolsep}{8pt}\begin{pmatrix} U & 0 \\ 0 & \bm{I}_{j} \end{pmatrix}},\quad
		V^{\dagger} &\to V_{\text{ext}}^{\dagger} = {\renewcommand{\arraystretch}{1}\setlength{\arraycolsep}{8pt}\begin{pmatrix} V^{\dagger} & 0 \\ 0 & \bm{I}_{j} \end{pmatrix}},
	\end{aligned}
\end{equation}
and
\begin{widetext}
	\begin{equation}
		\Sigma_{\text{ext}} = 
		{\renewcommand{\arraystretch}{1}\setlength{\arraycolsep}{8pt}
			\begin{pmatrix}
			\bm{I}_{k} & & & & & & & \\
			& \sigma_1 & \cdots & & & \sqrt{1-\sigma_1^2} & & \\
			& & \ddots & & & \vdots & \ddots & \\
			& & & \sigma_j &\cdots&  & &\sqrt{1-\sigma_j^2}\\
			& & & & \bm{0}_{N-j-k} & & & \\
			& & & & & 0 & & \\
			& & & & & & \ddots & \\
			& & & & & & & 0
		\end{pmatrix}}.
	\end{equation}
\end{widetext}
The non-unitary transformations $S_i$ are embedded in $\Sigma_{\text{ext}}$: matrix elements of $\Sigma_{\text{ext}}$ at $(k+i,k+i)$, $(k+i,k+i+j)$, $(k+i+j,k+i)$ and $(k+i+j,k+i+j)$ are elements of $S_i$, while other elements are from $\Sigma$. By doing so, the singular values of $\Sigma_{\text{ext}}$ are $\sigma\in\{0,1\}$, which can be obtained by performing SVD of $\Sigma_{\text{ext}}$ as
\begin{widetext}
	\begin{equation}
		\begin{aligned}
			\Sigma_{\text{ext}} = & I_{N+j}\cdot
			\text{diag}\left(\bm{I}_k,1_1,\cdots,1_j,\bm{0}_{N-j-k},0_1,\cdots,0_j\right)\\
			& \cdot
			{\renewcommand{\arraystretch}{1}\setlength{\arraycolsep}{8pt}
				\begin{pmatrix}
					I & & & & & & & \\
					& \sigma_1 & & & & \sqrt{1-\sigma_1^2} & & \\
					& & \ddots & & & & \ddots & \\
					& & & \sigma_j & & & & \sqrt{1-\sigma_j^2} \\
					& & & & I & & & \\
					& \sqrt{1-\sigma_1^2} & & & & -\sigma_1 & & \\
					& & \ddots & & & & \ddots & \\
					& & & \sqrt{1-\sigma_j^2} & & & & -\sigma_j
			\end{pmatrix}}.
		\end{aligned}
	\end{equation}
\end{widetext}
The enlarged scattering matrix is thus $M_{\text{ext}}=U_{\text{ext}}\Sigma_{\text{ext}}V_{\text{ext}}^{\dagger}$. 
Since $U_{\text{ext}}$ and $V_{\text{ext}}^{\dagger}$ are unitary matrices, singular values of $M_{\text{ext}}$ are determined by $\Sigma_{\text{ext}}$, which is therefore also $\in\{0,1\}$. Then the extended input Hilbert space $\mathcal{H}_{\text{ext}}^{\text{in}}$ can also be decomposed into several DPLSs. In this way, our theory of DPLS can also be applied. Finally, by tracing out the ancilla modes, output optical field can be obtained. 

We illustrate the above method by the following example. Consider a scattering matrix and its SVD
\begin{equation}
	\begin{aligned}
		M &= {\renewcommand{\arraystretch}{0.8}\setlength{\arraycolsep}{8pt}
			\begin{pmatrix} \frac{3}{4} & \frac{1}{4} \\[4pt] \frac{1}{4} & \frac{3}{4} \end{pmatrix}} \\
		&= \frac{1}{\sqrt{2}}{\renewcommand{\arraystretch}{0.8}\setlength{\arraycolsep}{8pt} 
			\begin{pmatrix} 1 & -1 \\[4pt] 1 & 1 \end{pmatrix}
			\begin{pmatrix} 1 & 0 \\[4pt] 0 & \frac{1}{2} \end{pmatrix}
			\frac{1}{\sqrt{2}} \begin{pmatrix} 1 & 1 \\[4pt] -1 & 1 \end{pmatrix}} \\
		&= U \Sigma V^{\dagger}
	\end{aligned}
\end{equation}
There is only one singular value $\sigma_2=\frac{1}{2}\in [0,1]$, so we just need to introduce one ancilla input mode $\hat{g}$ in vacuum state with its corresponding output mode $\hat{g}^{\text{out}}$, and a non-unitary transformation $S={\renewcommand{\arraystretch}{0.8}\setlength{\arraycolsep}{8pt}\begin{pmatrix}
		\frac{1}{2} & \frac{\sqrt{3}}{2} \\
		0 & 0
\end{pmatrix}}$, so that the scattering matrix is extended as 
\begin{equation}
	\begin{aligned}
		M_{\text{ext}} &= U_{\text{ext}} \Sigma_{\text{ext}} V_{\text{ext}}^{\dagger} \\
		&= {\renewcommand{\arraystretch}{0.8}\setlength{\arraycolsep}{8pt}
			\begin{pmatrix}
				\frac{1}{\sqrt{2}} & -\frac{1}{\sqrt{2}} & 0 \\[4pt]
				\frac{1}{\sqrt{2}} & \frac{1}{\sqrt{2}} & 0 \\[4pt]
				0 & 0 & 1
			\end{pmatrix}
			\begin{pmatrix}
				1 & 0 & 0 \\[4pt]
				0 & \frac{1}{2} & \frac{\sqrt{3}}{2} \\[4pt]
				0 & 0 & 0
			\end{pmatrix}
			\begin{pmatrix}
				\frac{1}{\sqrt{2}} & \frac{1}{\sqrt{2}} & 0 \\[4pt]
				-\frac{1}{\sqrt{2}} & \frac{1}{\sqrt{2}} & 0 \\[4pt]
				0 & 0 & 1
		\end{pmatrix}} \\[8pt]
		&= {\renewcommand{\arraystretch}{0.8}\setlength{\arraycolsep}{8pt}
			\begin{pmatrix}
				\frac{3}{4} & \frac{1}{4} & -\frac{\sqrt{6}}{4} \\[4pt]
				\frac{1}{4} & \frac{3}{4} & \frac{\sqrt{6}}{4} \\[4pt]
				0 & 0 & 0
		\end{pmatrix}}.
	\end{aligned}
\end{equation}
Its SVD gives 
\begin{equation}
	\begin{aligned}
		M_{\text{ext}} &= U_{\text{ext}}' \Sigma_{\text{ext}}' \left(V'_{\text{ext}}\right)^{\dagger} \\
		&= {\renewcommand{\arraystretch}{0.8}\setlength{\arraycolsep}{8pt}
			\begin{pmatrix}
				-\frac{3}{\sqrt{10}} & \frac{1}{\sqrt{10}} & 0 \\[6pt]
				\frac{1}{\sqrt{10}} & \frac{3}{\sqrt{10}} & 0 \\[6pt]
				0 & 0 & 1
			\end{pmatrix}
			\begin{pmatrix}
				1 & 0 & 0 \\[4pt]
				0 & 1 & 0 \\[4pt]
				0 & 0 & 0
		\end{pmatrix}}\cdot\\
		&\quad
		{\renewcommand{\arraystretch}{0.8}\setlength{\arraycolsep}{8pt}
			\begin{pmatrix}
				-\sqrt{\frac{2}{5}} & 0 & \sqrt{\frac{3}{5}} \\[10pt]
				\frac{3\sqrt{10}}{20} & \frac{\sqrt{10}}{4} & \frac{\sqrt{15}}{10} \\[10pt]
				\frac{\sqrt{6}}{4} & -\frac{\sqrt{6}}{4} & \frac{1}{2}
		\end{pmatrix}}.
	\end{aligned}
\end{equation}
Consequently, when one photon is input into the system, there are two relevant DPLSs
\begin{equation}
	\begin{aligned}
		\mathcal{H}_0^{\text{in}} = & \operatorname{Span} \Biggl\{  
		\left( -\sqrt{\frac{2}{5}} \hat{a}_1^{\dagger} + \sqrt{\frac{3}{5}} \hat{g}^{\dagger} \right) |0\rangle, \\
		&\qquad\quad \frac{3\sqrt{10} \hat{a}_1^{\dagger} + 5\sqrt{10} \hat{a}_2^{\dagger} + 2\sqrt{15} \hat{g}^{\dagger}}{20} |0\rangle \Biggr\}, \\[8pt]
		\mathcal{H}_1^{\text{in}} =& \operatorname{Span} \left\{ \frac{\sqrt{6} \hat{a}_1^{\dagger} - \sqrt{6} \hat{a}_2^{\dagger} + 2\hat{g}^{\dagger}}{4} |0\rangle \right\}.
	\end{aligned}
\end{equation}
Take $\vert 100\rangle $ input state as an example. It can be projected onto DPLSs as
\begin{equation}
	\vert 100\rangle=\frac{\sqrt{10}}{4}\vert \psi_0\rangle+\frac{\sqrt{6}}{4}\vert\psi_1\rangle.
\end{equation}
Projections $\vert\psi_0\rangle$ in $\mathcal{H}_0^{\text{in}}$ and $\vert\psi_1\rangle$ in $\mathcal{H}_1^{\text{in}}$ are respectively 
\begin{equation}
	\begin{aligned}
		|\psi_0\rangle &= -\frac{4}{5} \left( -\sqrt{\frac{2}{5}} \hat{a}_1^\dagger + \sqrt{\frac{3}{5}} \hat{g}^\dagger \right) |0\rangle \\
		& \quad+ \frac{3}{5} \left( \frac{3\sqrt{10} \hat{a}_1^\dagger + 5\sqrt{10} \hat{a}_2^\dagger + 2\sqrt{15} \hat{g}^\dagger}{20} \right) |0\rangle, \\[8pt]
		|\psi_1\rangle &= \frac{\sqrt{6} \hat{a}_1^\dagger - \sqrt{6} \hat{a}_2^\dagger + 2\hat{g}^\dagger}{4} |0\rangle.
	\end{aligned}
\end{equation}
According to Eq. (\ref{operator M}), the two projections then evolve into
\begin{equation}
	\begin{aligned}
		|\psi_{0,\text{out}}\rangle &= \hat{S}_{U'_{\text{ext}}}\hat{S}_{\left(V'_{\text{ext}}\right)^\dagger}\vert\psi_0\rangle\\
		& = \frac{1}{\sqrt{10}} \left( 3|100\rangle + |010\rangle \right), \\[4pt]
		|\psi_{1,\text{out}}\rangle &=\hat{S}_{U'_{\text{ext}}}\hat{g}\hat{S}_{\left(V'_{\text{ext}}\right)^\dagger}\vert\psi_0\rangle\\
		& = |000\rangle.
	\end{aligned}
\end{equation}
Output state with ancilla mode $\hat{g}$ is the statistical mixture of $|\psi_{0,\text{out}}\rangle$ and $|\psi_{1,\text{out}}\rangle$, therefore, 
\begin{equation}
	\begin{aligned}
		\hat{\rho}_{\text{out},\hat{g}} = & \frac{5}{8} \left( \frac{3}{\sqrt{10}} |100\rangle + \frac{1}{\sqrt{10}} |010\rangle \right) \left( \frac{3}{\sqrt{10}} \langle 100| + \frac{1}{\sqrt{10}} \langle 010| \right)\\
		& + \frac{3}{8} |000\rangle \langle 000|.
	\end{aligned}
\end{equation}
Tracing out the ancilla mode $\hat{g}$, the output two-mode optical state is
\begin{equation}
	\begin{aligned}
		\hat{\rho}_{\text{out}} = & \frac{5}{8} \left( \frac{3}{\sqrt{10}} |10\rangle + \frac{1}{\sqrt{10}} |01\rangle \right) \left( \frac{3}{\sqrt{10}} \langle 10| + \frac{1}{\sqrt{10}} \langle 01| \right) \\
		& + \frac{3}{8} |00\rangle \langle 00|.
	\end{aligned}
	\label{two-mode optical output}
\end{equation}
So our theory can also be applied to general lossy linear systems with singular values between $0$ and $1$.

\section{Decomposition of the $W$-state-generation system into beam splitters and phase shifters}\label{T&D}
In this section, we decompose the system for robust $W$-state generation into beam splitters and phase shifers, which can be easily realized through optical elements. According to the matrix decomposition in \cite{41_reck1994experimental}, $U$ and $V^{\dagger}$ in Eq. (\ref{lossy system for W generation}) can be further decomposed as Eq. (\ref{UVSigma for W generation}).
Through calculation, their detailed forms are 
\begin{equation}
	\begin{aligned}
		V^\dagger = 
		&\begin{pmatrix}
			\frac{1}{\sqrt{2}} & -\frac{1}{\sqrt{2}} & 0 \\[6pt]
			\frac{1}{\sqrt{3}} & \frac{1}{\sqrt{3}} & \frac{1}{\sqrt{3}} \\[6pt]
			\frac{1}{\sqrt{6}} & \frac{1}{\sqrt{6}} & -\frac{\sqrt{2}}{\sqrt{3}}
		\end{pmatrix} \\[10pt]
		= &\begin{pmatrix}
			\sqrt{\frac{3}{5}} & -\sqrt{\frac{2}{5}} & 0 \\[6pt]
			\sqrt{\frac{2}{5}} & \sqrt{\frac{3}{5}} & 0 \\[6pt]
			0 & 0 & 1
		\end{pmatrix}
		\cdot
		\begin{pmatrix}
			\sqrt{\frac{5}{6}} & 0 & \sqrt{\frac{1}{6}} \\[6pt]
			0 & 1 & 0 \\[6pt]
			-\sqrt{\frac{1}{6}} & 0 & \sqrt{\frac{5}{6}}
		\end{pmatrix} \\[10pt]
		\cdot &
		\begin{pmatrix}
			1 & 0 & 0 \\[4pt]
			0 & \frac{2}{\sqrt{5}} & \frac{1}{\sqrt{5}} \\[4pt]
			0 & -\frac{1}{\sqrt{5}} & \frac{2}{\sqrt{5}}
		\end{pmatrix}
		\cdot
		\begin{pmatrix}
			1 & \hspace{0.3cm}0\hspace{0.3cm} & 0 \\[4pt]
			0 & \hspace{0.3cm}1 \hspace{0.3cm}& 0 \\[4pt]
			0 & \hspace{0.3cm}0 \hspace{0.3cm}& -1
		\end{pmatrix} \\[10pt]
		= & T_{V,12} \, T_{V,13} \, T_{V,23} \, D_V,
	\end{aligned}
	\label{Vdag decomposition}
\end{equation}
and
\begin{align}
	U = 
	&\begin{pmatrix}
		-\frac{1}{\sqrt{3}} & \frac{1}{\sqrt{2}} & \frac{1}{\sqrt{6}} \\[6pt]
		-\frac{1}{\sqrt{3}} & -\frac{1}{\sqrt{2}} & \frac{1}{\sqrt{6}} \\[6pt]
		-\frac{1}{\sqrt{3}} & 0 & -\frac{\sqrt{2}}{\sqrt{3}}
	\end{pmatrix}\notag \\[10pt]
	=& \begin{pmatrix}
		\sqrt{\frac{2}{2}} & -\sqrt{\frac{2}{2}} & 0 \\[6pt]
		\sqrt{\frac{2}{2}} & \sqrt{\frac{2}{2}} & 0 \\[6pt]
		0 & 0 & 1
	\end{pmatrix}
	\cdot
	\begin{pmatrix}
		\sqrt{\frac{2}{3}} & 0 & -\sqrt{\frac{1}{3}} \\[6pt]
		0 & 1 & 0 \\[6pt]
		\sqrt{\frac{1}{3}} & 0 & \sqrt{\frac{2}{3}}
	\end{pmatrix} \\[10pt]
	\cdot&
	\begin{pmatrix}
		-1 & 0 & 0 \\[4pt]
		0 & -1 & 0 \\[4pt]
		0 & 0 & -1
	\end{pmatrix}\notag \\[10pt]
	=& T_{U,12} \, T_{U,13} \, D_U.\notag
\end{align}
Among them, $T$s can be constructed by two-mode beam splitters along with uncoupled lossless modes. For example, according to Eq. (\ref{Vdag decomposition}), $T_{V,12}$ can be constructed by a two mode beam splitter 
\begin{equation*}
	t_{V,12} = 
	\begin{pmatrix}
		\sqrt{3/5} & -\sqrt{2/5} \\
		\sqrt{2/5} & \sqrt{3/5}
	\end{pmatrix}
\end{equation*}
along with an uncoupled lossless mode. While $D$s are simply phase shifters changing the phase of each input mode. Therefore, the whole system for robust $W$-state generation can be implemented as in Fig. (\ref{fig:4}).

\end{document}